\documentclass[article,shortnames]{jss}

\usepackage{orcidlink,thumbpdf,lmodern}

\usepackage{framed}

\usepackage{stfloats}
\usepackage{amsthm,amsmath,amsfonts,amssymb}
\usepackage{mathrsfs}
\usepackage{bm}
\usepackage{epsfig}
\usepackage{multicol,multirow}
\usepackage{tabularx,ragged2e,booktabs,rotating}
\usepackage{makecell}
\usepackage{xcolor}
\usepackage{booktabs}
\usepackage[utf8]{inputenc}
\usepackage[T1]{fontenc}
\usepackage{textcomp}
\usepackage{hyperref}

\newcommand{\calY}{\mathcal{Y}}
\newcommand{\calL}{\mathcal{L}}
\newcommand{\calK}{\mathcal{K}}
\newcommand{\calC}{\mathcal{C}}

\newcommand{\iidsim}{\stackrel{\text{i.i.d.}}{\sim}}
\newcommand{\bOmega}{\bm{\Omega}}
\newcommand{\bSigma}{\bm{\Sigma}}

\newcommand{\hbSigma}{\hat{\bSigma}}
\newcommand{\by}{\bm{y}}

\newcommand{\bZ}{\bm{Z}}

\newcommand{\bB}{\bm{B}}
\newcommand{\bA}{\bm{A}}

\newcommand{\barby}{\bar{\by}}

\global\long\def\E{\operatorname{E}}
\global\long\def\COV{\operatorname{Cov}}
\global\long\def\tr{\operatorname{tr}}

\usepackage{soul,xcolor,cancel}
\usepackage{color}

\usepackage{soul,xcolor,cancel}

\setstcolor{red}
\newif\ifApproveEdit
\ApproveEditfalse
\ifApproveEdit

\newcommand{\del}[1]{\iffalse{#1}\fi}
\else

\newcommand\del[2][red]{\setbox0=\hbox{$#2$}%
	\rlap{\raisebox{.45\ht0}{\textcolor{#1}{\rule{\wd0}{1pt}}}}#2}
\fi

\newtheorem{definition}{Definition}

\author{Pengfei Wang\\National Institute \\of Education, \\Nanyang Technological \\University
   \And Tianming Zhu~\orcidlink{0000-0003-0798-6688}\\National Institute \\of Education, \\Nanyang Technological \\University
   \And Jin-Ting Zhang\\National University \\of Singapore}
\Plainauthor{Pengfei Wang, Tianming Zhu,  Jin-Ting Zhang}

\title{\pkg{HDNRA}: An \proglang{R} package for HDLSS location testing with normal-reference approaches}
\Plaintitle{HDNRA: An R package for high-dimensional location testing with normal-reference approaches}
\Shorttitle{\pkg{HDNRA}: Normal-reference approaches for high-dimensional data in \proglang{R}}

\Abstract{
The challenge of location testing for high-dimensional data in statistical inference is notable. Existing literature suggests various methods, many of which impose strong regularity conditions on underlying covariance matrices to ensure asymptotic normal distribution of test statistics, leading to difficulties in size control. To address this, a recent set of tests employing the normal-reference approach has been proposed. Moreover, the availability of tests for high-dimensional location testing in \proglang{R} packages implemented in \proglang{C++} is limited. This paper introduces the latest methods utilizing  normal-reference approaches to test the equality of mean vectors in high-dimensional samples with potentially different covariance matrices. We present an \proglang{R} package named \pkg{HDNRA} to illustrate the implementation of these tests, extending beyond the two-sample problem to encompass general linear hypothesis testing (GLHT). The package offers easy and user-friendly access to these tests, with its core implemented in \proglang{C++} using \pkg{Rcpp}, \pkg{OpenMP} and \pkg{RcppArmadillo} for efficient execution. Theoretical properties of these normal-reference tests are revisited, and examples based on real datasets using different tests are provided.
}

\Plainkeywords{R, high-dimensional data, normal-reference approach, two-sample problem, GLHT problem, C++, Rcpp, RcppArmadillo}

\Address{
  Pengfei Wang\\
  National Institute of Education\\
  Nanyang Technological University\\
  1 Nanyang Walk, 637616, Singapore\\
  E-mail: \email{nie23.wp8738@e.ntu.edu.sg}\\
  
  Tianming Zhu\\
  National Institute of Education\\
  Nanyang Technological University\\
  1 Nanyang Walk, 637616, Singapore\\
  E-mail: \email{tianming.zhu@nie.edu.sg}\\
  URL: \url{https://math.nie.edu.sg/about/staff/zhutianming.aspx}\\
  
  Jin-Ting Zhang\\
  Department of Statistics and Data Science\\
  National University of Singapore\\
  6 Science Drive 2, 117546, Singapore\\
  E-mail: \email{stazjt2020@nus.edu.sg}\\
  URL: \url{https://blog.nus.edu.sg/stazjt2020/}
}

\begin{document}


\section[Introduction]{Introduction} \label{sec:intro}

The exploration of high-dimensional low-sample-size (HDLSS) data, often referred to as "large $p$ and small $n$" ($n \ll p$) data, where $p$ signifies the number of variables and $n$ denotes the sample size, has been a focal point in both theoretical and applied research for several decades. The surge in interest can be attributed to technological advancements enabling the acquisition of an extensive array of variables in each analysis sample. This includes data from diverse fields such as genomic studies, biological research, financial analysis, satellite imaging, and related areas. A fundamental challenge in the analysis of high-dimensional data lies in comparing mean vectors across distinct populations.

Given $p$-variate random samples $\calY_i, i=1,\ldots,k$ from $k$ (where $k\geq 2$) independent populations, where $\calY_i=\{\bm{y}_{ij}=\left(y_{ij1}, \ldots, y_{ijp}\right)^{\top}, j=1,\ldots,n_i\}$, we assume that for each $i=1,\ldots,k,\; \bm{y}_{i1}, \ldots, \bm{y}_{i n_i}$ are independent and identically distributed (i.i.d.) with $\E(\bm{y}_{i1})=\bm{\mu}_i$ and $\COV(\bm{y}_{i1})=\bm{\Sigma}_{i}$. Our primary goal is to test the equality of the $k$ mean vectors:
\begin{equation}\label{hypothesis}
H_0: \bm{\mu}_1=\cdots=\bm{\mu}_k, \quad \mbox{vs.} \quad H_1: H_0  \text{ is not true.}
\end{equation}
When $p$ is fixed and much smaller than $n$ (where $n=n_1+\cdots+n_k$ is the total sample size), classical tests such as Hotelling's $T^2$-test by \cite{Hotelling_1992} and the Lawley--Hotelling trace test by \cite{dasgupta2005awley} can address this problem. However, as the dimension $p$ increases and approaches or surpasses the sample size $n$, a challenge arises because the $k$ sample covariance matrices, i.e., the natural estimators of the  $k$ covariance matrices $\bm{\Sigma}_1, \ldots, \bm{\Sigma}_k$ may become near singular or singular. Consequently, the classical tests may become less powerful or ill-defined in such cases. Recently, there has been significant interest in extending classical tests to the $p > n$ setting.


When $k=2$, the location testing problem in (\ref{hypothesis}) simplifies to the two-sample problem for high-dimensional data. Over the past six decades, numerous authors have proposed various alternatives. Some tests, as outlined in the works of \cite{Dempster_1958, Dempster_1960, bai1996effect, zhang2020simple}, assume equal covariance matrices, i.e., $\bm{\Sigma}_1 = \bm{\Sigma}_2$. When equal covariance matrices are not assumed, the problem is commonly known as the two-sample Behrens--Fisher (BF) problem. In recent decades, researchers have put forth a variety of tests to address this variation. Notable among them are some scale-invariant tests, such as those introduced by \cite{Srivastava_2008, Tony_Cai_2013, Srivastava_2013, Feng_2015, Gregory_2015, Dong_2016}. U-statistic-based tests have been explored by \cite{Chen_2010, Ahmad_2013, LongFeng_2015, Wang_2022}. Additionally, empirical likelihood ratio tests have been introduced by \cite{Zhou_2013, Kim_2015}. Simulation or permutation-based tests offer another approach, as proposed by \cite{Lee_2015, Wei_2016}. Random-projection and subspace-based tests have been suggested by \cite{lopes2011more, Thulin_2014, Ma_2015, Mondal_2015, Zhou_2015, Zhang_2016}. Furthermore, nonparametric tests have been explored by \cite{Li_2011, Ghosh_2015, Wang_2015}.

When $k>2$, the location testing problem in (\ref{hypothesis}) is well-known as the one-way multivariate analysis of variance (MANOVA) problem. Assuming $\bm{\Sigma}_1=\cdots=\bm{\Sigma}_k$, several significant contributions have been made. For instance, \cite{Fujikoshi_2004} derived the asymptotic normality of several classical MANOVA tests under a high-dimensional setting. Building on the work of \cite{bai1996effect}, \cite{Schott_2007} extended it to the one-way MANOVA problem for high-dimensional data. \cite{Srivastavaandtatsuya_2013} proposed a scale-invariant test for non-normal high-dimensional data. Investigating the asymptotics of Dempster's trace criterion, \cite{Nishiyama_2013} made noteworthy contributions. \cite{Cai_2014} introduced a linear transformation-based test that considers the dependence structure of the variables. Furthermore, \cite{Zhang_2017} explored the general linear hypothesis testing (GLHT) problem with a common covariance matrix, proposing an $L^2$-norm-based test.

In the context of the $k$-sample BF problem, where equal covariance matrices are not assumed, various approaches have been explored. \cite{Zhang_2009} proposed an approximate solution based on \cite{Bennett_1950}'s work. Other studies introduced novel techniques, such as \cite{Srivastava_2006}'s general linear hypothesis test under multivariate linear regression models for normal data, \cite{Zhou_2017}'s test using U-statistics, and investigations into the asymptotic normality by \cite{Yamada_2015} and \cite{Hu_2015}. Further contributions include \cite{Hyodo_2018}'s consideration of simultaneous confidence interval estimation for paired mean vectors, \cite{Chen_2019}'s powerful test for sparse and faint mean differences, and \cite{Watanabe_2020}'s study of the two-way MANOVA problem for high-dimensional data.

Recently, a novel set of tests based on the normal-reference approach has been introduced. The normal-reference approach, initially coined by \cite{Zhang_2021}, drew inspiration from the work of \cite{Zhang_2017}. This approach has been employed to formulate test statistics for addressing the high-dimensional location testing, encompassing both the two-sample problem and the general linear hypothesis testing (GLHT) problem. The primary advantage of the normal-reference approach lies in its departure from reliance on the limiting distribution of the proposed test statistic, which can be either normal or non-normal. Instead, it approximates the null distribution by utilizing its normal-reference distribution, derived by treating the high-dimensional data as if they were normally distributed. The resulting normal-reference tests demonstrate robust size control, irrespective of whether the high-dimensional data are nearly uncorrelated, moderately correlated, or highly correlated.
For the two-sample problem, $L^2$-norm-based tests were proposed by \cite{zhang2020simple} and \cite{Zhang_2021}, while scale-invariant tests were introduced by \cite{Zhang_2020} and \cite{zhang2023two}. Additionally,  \cite{zhang2022further} and \cite{zhang2022revisit} developed normal-reference tests with three-cumulant (3-c) matched $\chi^2$-approximation, complemented by an $F$-type test studied by \cite{zhu2022two}. Shifting focus to the GLHT problem, \cite{Zhang_2017} and \cite{zhang2022linear} proposed tests based on $L^2$-norm with Welch–Satterthwaite (W–S) $\chi^2$-approximation, while \cite{Zhu_2023} presented a scale-invariant test. Moreover, \cite{zhu2022linear} and \cite{Zhang_2022heteroscedastic} introduced normal-reference tests with 3-c matched $\chi^2$-approximation for the GLHT problem. In Section~\ref{sec:2}, we will delve into the concept of the normal-reference approach. Theoretical properties of the aforementioned normal-reference tests will be revisited and discussed in some detail. To the best of our knowledge, there is limited literature comprehensively presenting the latest research on high-dimensional location testing employing the normal-reference approach.

Several authors have contributed summarized works on the high-dimensional location testing. For instance, \cite{Hu_2016} reviewed naive testing methods for the mean vectors and covariance matrices of two high-dimensional populations. In a comprehensive review, \cite{paul2019review} not only summarized existing procedures but also proposed new ones, evaluated their size and power, and provided recommendations for the two-sample BF problem and its analogous problems for non-normal populations. Offering a selective overview, \cite{huang_2022} focused on the motivation behind testing procedures, insights into constructing test statistics, and connections and comparisons of different methods. In a categorization by \cite{Harrar_2022}, methods for $k$-sample problems are grouped into three categories—parametric, semi-parametric, and non-parametric—based on the hypothesis of interest and model assumptions. 


Many high-dimensional tests involve substantial computational complexity and require efficient expressions and code, especially for estimators of the asymptotic variances.  Several \proglang{R} packages for the location testing problem are available. For example, \pkg{stats} (\citealt{stats}) incorporates the Welch's $t$-test based on least squares estimators, while \pkg{asht} (\citealt{asht}) focuses on tests for the BF problem under normality.  Additionally, \pkg{WRS2} (\citealt{WRS2}) provides tests based on Yuen's approach utilizing trimmed sample means. \pkg{RobustBF} (\citealt{RobustBF}) computes the adaptive modified maximum likelihood test statistics and the corresponding $p$-values. Furthermore, the work of \cite{Harrar_2022} delves into several packages, including \pkg{SHT} (\citealt{SHT}), \pkg{highDmean} (\citealt{highDmean}), \pkg{HDtest} (\citealt{HDtest}), \pkg{ARHT} (\citealt{ARHT}), \pkg{highmean} (\citealt{highmean}), \pkg{MethylCapSig} (\citealt{MethylCapSig}), and \pkg{highD2pop} (\citealt{highD2pop}), analyzing and discussing their performance.  Among these, \pkg{nparcomp} (\citealt{nparcomp}), illustrated by \cite{konietschke2015nparcomp} includes a two-sample nonparametric studentized permutation test for paired data. Additionally, \pkg{NSM3} (\citealt{NSM3}) offers a non-parametric procedure test for the BF problem. \pkg{HDMT} (\citealt{HDMT}) is designed for high-dimensional mediation hypotheses, while  \pkg{TVMM} (\citealt{TVMM}), developed by \cite{de2023tvmm} is a statistical tool  that provides powerful multivariate statistical tests compared to traditional Hotelling $T^2$-test as well as the likelihood ratio test. 
There is also a \proglang{PYTHON} package named \pkg{hyppo} created by \cite{panda2019hyppo}, which addresses independence, two-sample, and $k$-sample testing problems. However, none of the available \proglang{R} packages for the high-dimensional location testing problem incorporates normal-reference tests, and only a limited number of packages encompass tests for the GLHT problem. Furthermore, it is worth noting that \pkg{HDtest}, \pkg{MethylCapSig}, and \pkg{highDpop} cannot be installed in the latest version of \proglang{R}. \pkg{HDtest} was built using the \proglang{Fortran} language, while \pkg{NSM3} and \pkg{highD2pop} were built using the \proglang{C} language. Among them, only \pkg{SHT} was built using \proglang{C++}.

In this article, we introduce the \proglang{R} package \pkg{HDNRA}, implemented in \proglang{C++} and built exclusively with standard libraries. The standalone \proglang{C++} version can be easily installed on any up-to-date platform. Leveraging the power of \pkg{Rcpp} (\citealt{eddelbuettel2011rcpp}) and \pkg{RcppArmadillo} (\citealt{RcppArmadillo}), our package seamlessly combines \proglang{R} and \proglang{C++}, resulting in a significant enhancement in the speed of its functions, as partially evidenced in Tables~\ref{tab:dataset_comparison1}--\ref{tab:dataset_comparison3} in Section~\ref{sec:5}.

\pkg{HDNRA} is designed for both academic research and real-data applications, serving as a fundamental toolbox for implementing a range of existing tests addressing the high-dimensional location testing problem in the \proglang{R} programming language. This package includes functions dedicated to executing various established tests for this problem, accompanied by two high-dimensional datasets. The functions not only provide the $p$-values generated by different tests but also include their corresponding test statistics and approximate  parameters. The package facilitates a quick and straightforward retrieval of $p$-values, enabling efficient detection of significant differences in mean vectors between different populations, and also is well-suited for some further contrast tests. In addition, by examining the estimated approximate degree of freedom, users can easily discern whether the null distribution of the test statistic is normal or nonnormal. This capability aids in determining the reliability of test results derived from those normal-approximation-based tests. Furthermore,  two extremely useful high-dimensional datasets are also suitable for use in the analyses conducted by other researchers working in the realm of high-dimensional data. Indeed, the utility of \pkg{HDNRA} extends beyond addressing the location testing alone. It can be effectively utilized in collaboration with other packages for high-dimensional data analysis (HDA). Since mean testing typically serves as the initial stage in  HDA, \pkg{HDNRA} contributes to informing subsequent analytical steps. All the code is open source and the development version of the package is hosted on GitHub
at \url{https://github.com/nie23wp8738/HDNRA}. Contributions are welcome both in terms of bug reports and feature enhancements, via the standard mechanism of GitHub issues and pull
requests. 


The remainder of this study is structured as follows. Section~\ref{sec:2} outlines the theoretical properties of the normal-reference approach and offers a summary of the corresponding normal-reference tests. Section~\ref{sec:3} introduces \pkg{HDNRA}, detailing its primary functions and underlying utility. In Section~\ref{sec:4}, we delineate the two datasets bundled within \pkg{HDNRA} and furnish a step-by-step guide on utilizing the package for both the two-sample problem and the GLHT problem, employing the two datasets. Section~\ref{sec:5} provides a comparative analysis with alternative packages.
We conclude with a summary and discussion in Section~\ref{sec:6}. Additional code is furnished in the Appendix.


\section{Normal-reference approaches and related tests}\label{sec:2}
\subsection{The normal-reference approach}

The normal-reference approach can be outlined as follows.  Suppose we have $k$ independent samples $\calY_i,i=1,\ldots,k$. Consider the test statistic $T_{n,p}(\calY_1,\ldots,\calY_k)$, with its distribution denoted as $\calL[T_{n,p}(\calY_1,\ldots,\calY_k)]$. By treating $\calY_i,i=1,\ldots,k$  as if they were normally distributed, designated as $\calY^*_i,i=1,\ldots,k$, we label $\calL[T_{n,p}(\calY^*_1,\ldots,\calY^*_k)]$ as the "normal-reference distribution" of $T_{n,p}(\calY_1,\ldots,\calY_k)$. In many instances, obtaining the original distribution $\calL[T_{n,p}(\calY_1,\ldots,\calY_k)]$ is challenging. However, acquiring the normal-reference distribution $\calL[T_{n,p}(\calY^*_1,\ldots,\calY^*_k)]$ is generally more manageable due to the normality assumption on $(\calY^*_1,\ldots,\calY^*_k)$. If we can demonstrate that under the null hypothesis and specific regularity conditions, $\calL[T_{n,p}(\calY_1,\ldots,\calY_k)]=\calL[T_{n,p}(\calY^*_1,\ldots,\calY^*_k)]$ asymptotically, it justifies approximating the original distribution $\calL[T_{n,p}(\calY_1,\ldots,\calY_k)]$ using its normal-reference distribution $\calL[T_{n,p}(\calY^*_1,\ldots,\calY^*_k)]$.
Often the latter can be well approximated using a two or three-cumulant matched $\chi^2$-approximation. The resulting test is then naturally termed a normal-reference test.

A normal-reference test is applicable when the test statistic  for the location testing problem (\ref{hypothesis}), i.e., $T_{n,p}(\calY_1,\ldots,\calY_k)$,  is  constructed  based on the $L^2$-norm of the differences between the sample means. It is often  easy to show  that the "normal-reference distribution" of $T_{n,p}(\calY_1,\ldots,\calY_k)$, i.e., $\calL[T_{n,p}(\calY^*_1,\ldots,\calY^*_k)]$ is equivalent to the distribution of a $\chi^2$-type mixture, denoted as $T_{n,p}^*$, for any fixed $n$ and $p$. Therefore, using $\calL(T_{n,p}^*)$ to approximate $\calL[T_{n,p}(\calY_1,\ldots,\calY_k)]$ is reasonable. Throughout this paper, let $\chi_v^2$ denote a central $\chi^2$-distribution with $v$ degrees of freedom. Then, under the null hypothesis, $T_{n,p}^*$ can generally be expressed as
\begin{equation}\label{chi}
T_{n,p}^*=\sum_{r=1}^p c_{n, p, r} A_r, \; A_r\iidsim \chi_{d_r}^2\;\text{ independent},
\end{equation}
where $c_{n, p, r}, r=1, \ldots,p$ are nonzero unknown real coefficients, and $d_r$'s are known integers representing the degrees of freedom of the chi-squared random variables. It is worth noting that estimating the coefficients $c_{n, p, r}, r=1, \ldots,p$ of $T_{n,p}^*$ \eqref{chi} can be challenging, especially when $p >n$. The approach by \cite{imhof1961computing} to compute the exact distribution of a $\chi^2$-type mixture with a few known coefficients cannot be directly applied to find the distribution of $T_{n,p}^*$ \eqref{chi}. \cite[Sec.4.3.2]{zhang2013analysis} introduced the following two-cumulant or three-cumulant matched $\chi^2$-approximation methods to approximate the distribution of $T_{n,p}^*$ \eqref{chi}.

\subsubsection*{Two-cumulant (2-c) matched $\chi^2$-approximation} \label{2c.sec}

When the coefficients of $T_{n,p}^*$ are all non-negative, the distribution of $T_{n,p}^*$ can be well approximated by  the two-cumulant (2-c) matched $\chi^2$-approximation,  also known as the Welch--Satterthwaite (W--S) $\chi^2$-approximation (\citealt{satterthwaite1946approximate, welch1947generalization}), or the Box $\chi^2$-approximation (\citealt{box1954some}). The key idea is to approximate $\calL(T_{n,p}^*)$ by $\calL(R_2)$ with $R_2 \stackrel{d}{=} \beta \chi_d^2$,
where $\stackrel{d}{=}$ denotes equality in distribution. The unknown parameters $\beta$ and $d$  can be determined via matching the  first two cumulants (means and variances)  of $T_{n,p}^*$  and $R_2$. We call $d$ the approximate degrees of freedom of the W--S $\chi^2$-approximation. 
By matching the first two cumulants of $T_{n,p}^*$ and $R_2$, we have
\begin{equation}\label{2c}
\beta=\frac{\mathcal{K}_2(T_{n,p}^*)}{2\mathcal{K}_1(T_{n,p}^*)}, \;\text { and }\; d=\frac{2\mathcal{K}_1^2(T_{n,p}^*)}{\mathcal{K}_2(T_{n,p}^*)},
\end{equation}
where $\mathcal{K}_{\ell}(T_{n,p}^*)=2^{\ell-1}(\ell-1)!\sum_{r=1}^p c_{n,p,r}^{\ell} d_r$ denotes the $\ell$-th cumulant of $T_{n,p}^*$ (\citealt{zhang2005approximate}) for $\ell=1,2,\ldots.$ 

\subsubsection*{Three-cumulant (3-c) matched $\chi^2$-approximation}\label{3c.sec}

When some  coefficients of $T_{n,p}^*$ are negative and some  are positive, the distribution of $T_{n,p}^*$ cannot be  well approximated by the two-cumulant (2-c) matched $\chi^2$-approximation described above. Instead, its distribution can be well approximated by the three-cumulant (3-c) matched $\chi^2$-approximation (\citealt{zhang2005approximate}). The key idea  is to approximate $\calL(T_{n,p}^*)$ by $\calL(R_3)$ with $R_3 \stackrel{d}{=} \beta_0+\beta_1 \chi_{d^*}^2$,
where $\beta_0, \beta_1$, and $d^*$ are unknown parameters  determined via matching the first three cumulants (means, variances, and third central moments) of $T_{n,p}^*$ and $R_3$.   Matching the first three cumulants of $T_{n,p}^*$ and $R_3$ leads to
\begin{equation}\label{3c}
\beta_0=\calK_1(T_{n,p}^*)-\frac{2\calK_2^2(T_{n,p}^*)}{\calK_3(T_{n,p}^*)}, \;\; \beta_1=\frac{\mathcal{K}_3(T_{n,p}^*)}{4\mathcal{K}_2(T_{n,p}^*)}, \;\text { and }\; d^*=\frac{8\mathcal{K}_2^3(T_{n,p}^*)}{\mathcal{K}_3^2(T_{n,p}^*)}.
\end{equation}

Let $\hat{\beta},\hat{d},\hat{\beta}_0,\hat{\beta}_1$, and $\hat{d}^*$ be the ratio-consistent estimators of $\beta,d,\beta_0,\beta_1$, and $d^*$, respectively. Then the proposed normal-reference test $T_{n,p}$ which related to the relevant literature can be conducted via using the critical value $\hat{\beta} \chi_{\hat{d}}^2(\alpha)$ (2-c matched $\chi^2$-approximation) or $\hat{\beta}_0 + \hat{\beta}_1 \chi_{\hat{d}^*}^2(\alpha)$ (3-c matched $\chi^2$-approximation), or the $p$-value $\Prob(\chi_{\hat{d}}^2 \geq T_{n, p} / \hat{\beta})$ (2-c matched $\chi^2$-approximation) or  $\Prob[\chi_{\hat{d}^*}^2 \geq (T_{n, p}-\hat{\beta}_0) / \hat{\beta}_1]$ (3-c matched $\chi^2$-approximation). In practical applications,  $\hat{\beta}$, $\hat{d}$, $\hat{\beta}_0$, $\hat{\beta}_1$, and $\hat{d}^*$ must be derived from the available data. In the interest of brevity, the specific methodologies for their estimation are not expounded upon in this paper. For comprehensive details, readers are directed to the pertinent references that will be cited later in our discussion of the related tests.

Before revisiting the normal-reference tests and referencing Definition 1 from \cite{huang_2022}, we provide the following definition.
\begin{definition}\label{ICM1}
A random sample $\calY=\{\by_{1},\ldots,\by_{n}\}$ is said to be generated from an independent component model (ICM) if we can write $\bm{y}_{i}=\bm{\mu}+\bm{\Gamma} \bm{z}_{i}, i=1, \ldots, n$, where $\bm{\Gamma}$ is a $p \times m$ matrix for some $m \geq p$ such that $\bm{\Gamma} \bm{\Gamma}^{\top}=\bm{\Sigma}$ and $\bm{z}_{i}=(z_{i1},\ldots,z_{im})^\top,i=1,\ldots,n$ are $m$-dimensional random vectors with i.i.d. elements $z_{ij}$'s with  $\E(z_{ij})=0,  \E(z_{ij}^2)=1$,  and $\E(z_{ij}^4)<\infty$.
\end{definition}
This model maintains that the observations $\by_i$ are linearly generated by $m$-variate $\bm{z}_i$ whose components are largely white noise and  $z_{i j}$  has independent structure. Then any samples generated from the ICM satisfy the conditions of \cite{bai1996effect} and \cite{Chen_2010}. Throughout this paper, we assume that all the independent samples are generated from the ICM distributions, and the sample sizes are balanced, i.e.,  as $n \rightarrow  \infty$, we have $n_i / n \rightarrow \tau_i \in(0,1), i=1,\ldots,k$. This assumption is regularity for the high-dimensional location testing  problem  which ensures that the $k$ sample sizes $n_i, i=1, \ldots, k$ tend to infinity proportionally. Further, let 
\[
\bar{\by}=n_i^{-1}\sum_{j=1}^{n_i}\by_{ij},\;\mbox{ and }\; \hat{\bm{\Sigma}}_i=\left(n_i-1\right)^{-1} \sum_{j=1}^{n_i}\left(\bm{y}_{i j}-\overline{\bm{y}}_i\right)\left(\bm{y}_{i j}-\overline{\bm{y}}_i\right)^{\top}, i=1,\ldots,k,
\]
be the sample mean vectors and sample covariance matrices, respectively, and  $\lambda_{ir},r=1,\ldots,p$ be the eigenvalues of $\bSigma_i,i=1,\ldots,k$ in descending order.

\subsection{Tests for the two-sample problem}\label{twosample}

For the hypothesis testing problem \eqref{hypothesis}, when $k=2$, if $p$ is fixed and $p < n$, by assuming $\bSigma_1=\bSigma_2=\bSigma$,  \cite{anderson2009introduction} stated that the classical Hotelling's $T^2$-test (\citealt{Hotelling_1992}) is the most powerful invariant test whose test statistic is defined as:
\begin{equation}\label{T_H}
T_{\mathrm{H}}=\frac{n_1n_2}{n}(\overline{\bm{y}}_1-\overline{\bm{y}}_2)^{\top} \hat{\bm{\Sigma}}^{-1}\left(\overline{\bm{y}}_1-\overline{\bm{y}}_2\right),
\end{equation}
where 
\begin{equation}\label{pSigma.sec2}
\hat{\bm{\Sigma}}=(n-2)^{-1}[(n_1-1)\hbSigma_1+(n_2-1)\hbSigma_2], 
\end{equation}
is the pooled sample covariance matrix. Without assuming  $\bSigma_1=\bSigma_2$, we can use the classical Wald-type test statistic for the two-sample BF problem:
\begin{equation}\label{T_W}
T_{\mathrm{W}}=\frac{n_1n_2}{n}(\overline{\bm{y}}_1-\overline{\bm{y}}_2)^{\top} \hat{\bm{\Sigma}}_n^{-1}\left(\overline{\bm{y}}_1-\overline{\bm{y}}_2\right),
\end{equation}
where:
\begin{equation}\label{hbOmegan}
    \hat{\bm{\Sigma}}_n = \frac{n_2}{n}\hbSigma_1+\frac{n_1}{n}\hbSigma_2,
\end{equation}
is the usual unbiased estimator of the covariance matrix $\bm{\Sigma}_n$ of $\sqrt{n_1n_2/n}(\overline{\bm{y}}_1-\overline{\bm{y}}_2)$, i.e.,
\begin{equation}\label{bOmegan}
     \bm{\Sigma}_n=\COV\Big[\sqrt{n_1n_2/n}(\overline{\bm{y}}_1-\overline{\bm{y}}_2)\Big]=\frac{n_2}{n}\bm{\Sigma}_1 +\frac{n_1}{n}\bm{\Sigma}_2.
\end{equation}
Throughout this paper, let  $\operatorname{tr}(\bm{A})$ denote  the trace of a matrix $\bm{A}$. To describe various testing procedures for the two-sample problem for high-dimensional data, we list the following assumptions:

\begin{center}
    Two-Sample Problem Assumptions (TA)
\end{center}
\begin{enumerate}
\item As $n,p\rightarrow \infty$, any of the following conditions holds
\begin{equation}\label{norconds.sec2}
	\begin{array}{ll}
		\lambda_{n,p,max}^2 = o\{\operatorname{tr}(\bm{\Sigma}_n^2)\}, &\mbox{\cite{bai1996effect}}, \\
		\operatorname{tr}(\bm{\Sigma}_n^4) = o\{\operatorname{tr}^2(\bm{\Sigma}_n^2)\},   &\mbox{\cite{Chen_2010}},\\
		\operatorname{tr}(\bm{\Sigma}_n^\ell)/p^2 \rightarrow a_\ell  \in (0,\infty), \ell=1,2,3,  &\mbox{\cite{Srivastava_2008}},
	\end{array}
	\end{equation}
	where $\lambda_{n,p,max}$ denotes the largest eigenvalue of $\bm{\Sigma}_n$ which is defined in (\ref{bOmegan}).
\item Let $\lambda_{n,p,r},r=1,\ldots,p$ be the eigenvalues of $\bm{\Sigma}_n$ (\ref{bOmegan}) in descending order. Set $\rho_{n, p, r}= \lambda_{n,p,r}/\sqrt{\operatorname{tr}(\bm{\Sigma}_n^2)},r=1,\ldots,p$ which are the eigenvalues of $\bm{\Sigma}_n/\sqrt{\operatorname{tr}(\bm{\Sigma}_n^2)}$ in descending order. There exist real numbers $\rho_{r}, r=1,2, \ldots$, such that $\lim _{p \rightarrow \infty} \rho_{n, p, r}=\rho_{r}, r=1,2, \ldots$, uniformly and $\lim _{p \rightarrow \infty} \sum_{r=1}^p \rho_{n, p, r}=\sum_{r=1}^{\infty} \rho_{r}<\infty$.
\item Let $\gamma_{n,p,r},r=1,\ldots,p$ be the eigenvalues of $\bm{R}_n=\bm{D}_n^{-1 / 2} \bm{\Sigma}_n\bm{D}_n^{-1 / 2}$ in descending order with $\bm{D}_n=\operatorname{diag}(\bm{\Sigma}_n)$. Set $\nu_{n, p, r}= \gamma_{n,p,r}/\sqrt{\operatorname{tr}(\bm{R}_n^2)},r=1,\ldots,p$ which are the eigenvalues of $\bm{R}_n/\sqrt{\operatorname{tr}(\bm{R}_n^2)}$ in descending order. There exist real numbers $\nu_{r}, r=1,2, \ldots$, such that $\lim _{p \rightarrow \infty} \nu_{n, p, r}=\nu_{r}, r=1,2, \ldots$, uniformly and $\lim _{p \rightarrow \infty} \sum_{r=1}^p \nu_{n, p, r}=\sum_{r=1}^{\infty} \nu_{r}<\infty$.
\item There exist two constants $c_1$ and $c_2$ such that $0<c_1 \leq \min _{1 \leq r \leq p} \sigma_{i,rr} \leq \max_{1 \leq r \leq p} \sigma_{i, rr} \leq c_2<\infty$ for all $p$ and $i=1,2$, where $\sigma_{i, rr},r=1\ldots,p;\;i=1,2$ are the diagonal entries of $\bSigma_n$.
\end{enumerate}

Note that the proof of Theorem 3 in \cite{zhang2020simple} showed that the first two  conditions in TA1 (\ref{norconds.sec2}) are equivalent.


When we assume $\bSigma_1=\bSigma_2=\bSigma$, the resulting matrices are $\bm{\Sigma}_n=\bSigma$ and $\hat{\bm{\Sigma}}_n$ should be replaced with $\hbSigma$ (\ref{pSigma.sec2}). When $p>n$, both $\hbSigma_1$ and $\hbSigma_2$ become singular. Consequently, traditional methods like the Hotelling $T^2$-test $T_H$ (\ref{T_H}) and the classical Wald-type test $T_W$ (\ref{T_W}) are unsuitable for the high-dimensional context. A straightforward procedure is to substitute  $\hat{\bm{\Sigma}}$ with the identity matrix $\bm{I}_p$, forming a sum-of-squares-type test statistic, which is equivalent to be directly based on the $L^2$-norm of the sample mean differences.   \cite{bai1996effect} firstly proposed such a test statistic based on $\left\|\overline{\bm{y}}_1-\overline{\bm{y}}_2\right\|^2$  where and throughout $\|\bm{a}\|$ denotes the $L^2$-norm of a vector $\bm{a}$. It  is constructed as  an unbiased estimate of $\left\|\bm{\mu}_1-\bm{\mu}_2\right\|^2$ and  can be equivalently written as: 
\begin{equation}\label{T_BS}
T_{\mathrm{BS}}=\frac{n_1n_2}{n} \left\|\overline{\bm{y}}_1-\overline{\bm{y}}_2\right\|^2-\operatorname{tr}(\hat{\bm{\Sigma}}).
\end{equation}
Under TA1, if as $n, p \rightarrow \infty$, we have $p / n \rightarrow c \in(0, \infty)$, \cite{bai1996effect} derived the asymptotic normality of $T_{\mathrm{BS}}$ under $H_0$, and showed theoretically and with extensive simulation studies that their test has much higher power than $T_H$.

\cite{Chen_2010} first noted that some strong moment conditions in \cite{bai1996effect} are due to the terms $\sum_{j=1}^{n_i} \bm{y}_{i j}^{\top} \bm{y}_{i j}, i =1,2$, in the expansion of $\left\|\overline{\bm{y}}_1-\overline{\bm{y}}_2\right\|^2$. However, these two terms are not useful in the two-sample testing problem. They then proposed the following U-statistics-based test statistic:
\begin{equation}\label{T_CQ}
T_{\mathrm{CQ}}=\frac{\sum_{i \neq j}^{n_1} \bm{y}_{1 i}^{\top} \bm{y}_{1 j}}{n_1\left(n_1-1\right)}+\frac{\sum_{i \neq j}^{n_2} \bm{y}_{2 i}^{\top} \bm{y}_{2 j}}{n_2\left(n_2-1\right)}-2 \frac{\sum_{i=1}^{n_1} \sum_{j=1}^{n_2} \bm{y}_{1 i}^{\top} \bm{y}_{2 j}}{n_1 n_2}.
\end{equation}

Without assuming $\bSigma_1=\bSigma_2$, \cite{Chen_2010} established the asymptotic normality of $T_{\mathrm{CQ}}$ under the null hypothesis and TA1. 
Further,  \cite{Chen_2010} studied the asymptotic properties of $T_{\mathrm{CQ}}$ and derived its asymptotic power under more general settings and weaker technical conditions than those given by \cite{bai1996effect}. 

\subsubsection*{Normal-reference tests with 2-c matched $\chi^2$-approximation}\label{nrt2c.sec}

\cite{zhang2020simple} proposed the following test statistic:
\begin{equation}\label{T_ZGZC}
T_{\mathrm{ZZGZ}} = \frac{n_1n_2}{n} \|\bar{\bm{y}}_1 - \bar{\bm{y}}_2\|^2,
\end{equation}
under the equal covariance matrix assumption, and \cite{Zhang_2021} extended the work of  \cite{zhang2020simple} to the two-sample BF problem for high-dimensional data. Notice that  the null distributions of $T_{\mathrm{BS}}$, $T_{\mathrm{CQ}}$, and several $L^2$-norm-based test statistics (to be introduced) are approximated using normal distributions. Instead of relying on  a normal approximation, 
 \cite{Zhang_2021}  showed that under either TA1 or TA2 and the null hypothesis, $\calL[T_{\mathrm{ZZGZ}}(\calY_1,\calY_2)]=\calL[T_{\mathrm{ZZGZ}}(\calY_1^*,\calY_2^*)]$ asymptotically, and $\calL[T_{\mathrm{ZZGZ}}(\calY_1^*,\calY_2^*)]$ is the same as the distribution of the following   $\chi^2$-type mixture: 
$T_{\mathrm{ZZGZ}}^*=\sum_{r=1}^p \lambda_{n,p, r} A_r, \; A_r\iidsim \chi_{1}^2$,
where $\lambda_{n,p,r}, r=1,\ldots,p$ are defined in TA2.
 Hence it is justified to use $\calL(T_{\mathrm{ZZGZ}}^*)$ to approximate $\calL[T_{\mathrm{ZZGZ}}(\calY_1,\calY_2)]$. However, the coefficients of $T_{\mathrm{ZZGZ}}^*$, i.e.,  $\lambda_{n,p,r}, r=1,\ldots,p$ are unknown and difficult to be estimated consistently. Fortunately, since the coefficients are all non-negative, \cite{Zhang_2021} approximated $\calL(T_{\mathrm{ZZGZ}}^*)$ using  the W--S $\chi^2$-approximation as described in Section~\ref{2c.sec}. Therefore, their test can be conducted easily.

\cite{Zhang_2021} did not take the variation of $T_{\mathrm{ZZGZ}}$ into account, and certain simulation results presented in their study indicated that when the total sample size $n$ is small, the precision of size control for $T_{\mathrm{ZZGZ}}$ may be compromised. In addressing this issue, \cite{zhu2022two} accounted for the variability of $T_{\mathrm{ZZGZ}}$ and, following the construction of the classical $F$-type statistic in both univariate and multivariate data analysis, proposed the subsequent $F$-type test statistic:
\begin{equation}\label{T_ZWZ}
T_{\mathrm{ZWZ}}=\frac{T_{\mathrm{ZZGZ}}}{S_\mathrm{ZWZ}}=\frac{n_1n_2n^{-1}\|\bar{\bm{y}}_1-\bar{\bm{y}}_2\|^2}{\operatorname{tr}(\hat{\bm{\Sigma}}_n)},
\end{equation}
where $S_\mathrm{ZWZ}=S_\mathrm{ZWZ}(\calY_1,\calY_2)=\operatorname{tr}(\hat{\bm{\Sigma}}_n)$.
\cite{Zhang_2021} showed that $\calL[T_{\mathrm{ZZGZ}}(\calY_1^*,\calY_2^*)]=\calL(T_{\mathrm{ZZGZ}}^*)$, and it is also easy to show that $\calL[S_\mathrm{ZWZ}(\calY_1^*,\calY_2^*)]$
is the same as the distribution of the following $\chi^2$-type mixture:
$$
S_{ZWZ}^* \stackrel{d}{=} \frac{n_2}{n\left(n_1-1\right)} \sum_{r=1}^p \lambda_{1 r} B_{1 r}+\frac{n_1}{n\left(n_2-1\right)} \sum_{r=1}^p \lambda_{2 r} B_{2 r}, 
B_{1 r} \stackrel{\text { i.i.d. }}{\sim} \chi_{n_1-1}^2, B_{2 r} \stackrel{\text { i.i.d. }}{\sim} \chi_{n_2-1}^2.
$$
Note that under the Gaussian assumption, $T_{\mathrm{ZZGZ}}$ and $S_\mathrm{ZWZ}$ are independent. Therefore,  for any given $n$ and $p$,  $\calL[T_{\mathrm{ZWZ}}(\calY_1^*,\calY_2^*)]=\calL(T_{\mathrm{ZWZ}}^*)$ with 
\[
T_{\mathrm{ZWZ}}^*\stackrel{d}{=}  \frac{T^*_{ZZGZ}}{S^*_{ZWZ}}\stackrel{d}{=} \frac{\sum_{r=1}^p \lambda_{n, p, r} A_r}{\left[\left(n_1-1\right)^{-1} n_2 \sum_{r=1}^p \lambda_{1 r} B_{1 r}+\left(n_2-1\right)^{-1} n_1 \sum_{r=1}^p \lambda_{2 r} B_{2 r}\right] / n}.
\]
\cite{zhu2022two} showed that under TA1 or TA2 and the null hypothesis, $\calL[T_{\mathrm{ZWZ}}(\calY_1,\calY_2)]=\calL[T_{\mathrm{ZWZ}}(\calY^*_1,\calY^*_2)]$ asymptotically. It follows that it is reasonable  to  approximate $\calL[T_{\mathrm{ZWZ}}(\calY_1,\calY_2)]$ using $\calL(T_{\mathrm{ZWZ}}^*)$. Since both the numerator and denominator of $T_{\mathrm{ZWZ}}^*$, i.e., $T^*_{ZZGZ}$ and $S^*_{ZWZ}$,  are $\chi^2$-type mixtures with unknown non-negative coefficients, their distributions  can be well approximated by the distributions of $S_1=\beta_1 \chi_{d_1}^2$ and $S_2=\beta_2 \chi_{d_2}^2$, respectively, obtained  using the 2-c matched $\chi^2$-approximation as described in Section~\ref{2c.sec}, where $\beta_1, \beta_2, d_1$, and $d_2$ are the approximation parameters determined via matching the first two cumulants  of $T^*_{ZZGZ}$ and $S_1$, and $S^*_{ZWZ}$ and $S_2$,  respectively.    Notice that   $\operatorname{E}(T^*_{\mathrm{ZZGZ}})=\operatorname{E}(S^*_\mathrm{ZWZ})$, we have $\beta_1 d_1=\beta_2d_2$.  As a result, $\calL(T_{\mathrm{ZWZ}}^*)$ can be approximated by $\calL(F)$ with $F= S_1/S_2\stackrel{d}{=}(\chi_{d_1}^2/d_1)/(\chi_{d_2}^2/d_2)\sim F_{d_1,d_2}$, where $F_{d_1, d_2}$ denotes the usual $F$ distribution with $d_1$ and $d_2$ degrees of freedom. Let $\hat{d}_1$ and $\hat{d}_2$ be the ratio-consistent estimators of $d_1$ and $d_2$; see \cite{zhu2022two} for details.   Then for any nominal significance level $\alpha>0$, the proposed normal-reference  $F$-type test can be conducted via using the critical value $F_{\hat{d}_1, \hat{d}_2}(\alpha)$ or the $p$-value $\Prob(F_{\hat{d}_1, \hat{d}_2} \geq F_{n, p})$ where $F_{v_1, v_2}(\alpha)$ denotes the upper $100 \alpha$ percentile of $F_{v_1, v_2}$.

\subsubsection*{Normal-reference tests with 3-c matched $\chi^2$-approximation}\label{nrt3c.sec}

\cite{zhang2022revisit} revisited \cite{bai1996effect}'s test and employed the same test statistic as $T_{\mathrm{BS}}$ \eqref{T_BS}. Instead of using normal approximation to the test statistic as done by \cite{bai1996effect},  they showed that under either TA1 or TA2 and the null hypothesis, $\calL[T_{\mathrm{BS}}(\calY_1,\calY_2)]=\calL[T_{\mathrm{BS}}(\calY_1^*,\calY_2^*)]$ asymptotically, and $\calL[T_{\mathrm{BS}}(\calY_1^*,\calY_2^*)]$  is the same as $\calL(T_{\mathrm{BS}}^*)$ with
    $T_{\mathrm{BS}}^*=\sum_{r=1}^p \lambda_{p, r}[A_r-B_r /(n-2)]$, where $A_r \stackrel{\text { i.i.d. }}{\sim} \chi_1^2$, and $B_r \stackrel{\text { i.i.d. }}{\sim} \chi_{n-2}^2$
are independent and $\lambda_{p,r},r=1,\ldots,p$ are the eigenvalues of the common covariance matrix $\bSigma$. Since $T_{\mathrm{BS}}^*$ can be either normal or non-normal, it is not appropriate to use the normal approximation blindly. Rather it is justified to use $\calL(T_{\mathrm{BS}}^*)$ to approximate $\calL[T_{\mathrm{BS}}(\calY_1,\calY_2)]$. As the coefficients of $T_{\mathrm{BS}}^*$ can be both positive and negative, they employed the 3-c matched $\chi^2$-approximation as described in Section~\ref{3c.sec} to approximate $\calL(T_{\mathrm{BS}}^*)$ instead of the W--S $\chi^2$-approximation which was used in \cite{zhang2020simple} and \cite{Zhang_2021}.

To reduce the computation time associated with $T_{\mathrm{CQ}}$, \cite{zhang2022further}  equivalently re-write $T_{CQ}$ (\ref{T_CQ}) as: 
\begin{equation}\label{T_ZZ_CQ}
T_{\mathrm{ZZ}} =  \|\bar{\bm{y}}_1 - \bar{\bm{y}}_2\|^2-\frac{n}{n_1n_2}\operatorname{tr}(\hat{\bm{\Sigma}}_n),
\end{equation}
where $\hat{\bm{\Sigma}}_n$ is defined in (\ref{hbOmegan}).
They showed that under either TA1 or TA2 and the null hypothesis, $\calL[T_{\mathrm{ZZ}} (\calY_1,\calY_2)]=\calL[T_{\mathrm{ZZ}} (\calY_1^*,\calY_2^*)]$ asymptotically, and $\calL[T_{\mathrm{ZZ}}(\calY_1^*,\calY_2^*)]=\calL(T_{\mathrm{ZZ}}^*) $  with
$T_{ZZ}^*=n(n_1n_2)^{-1}\sum_{r=1}^p \lambda_{n, p, r} A_r-\{[n_1(n_1-1)]^{-1}\sum_{r=1}^p \lambda_{1 r} B_{1 r}+[n_2(n_2-1)]^{-1}\sum_{r=1}^p \lambda_{2 r} B_{2 r}\}$,
where $\lambda_{n,p,r}$'s are defined in TA2, while $A_r \stackrel{\text { i.i.d. }}{\sim} \chi_1^2$, $B_{1 r} \stackrel{\text { i.i.d. }}{\sim} \chi_{n_1-1}^2$, and $B_{2 r} \stackrel{\text { i.i.d. }}{\sim} \chi_{n_2-1}^2$ are mutually independent.  Hence it is justified to use $\calL(T_{ZZ}^*)$ to approximate $\calL[T_{ZZ}(\calY_1,\calY_2)]$. Further, $\calL(T_{ZZ}^*)$ can be approximated by employing the 3-c matched $\chi^2$-approximation as described in Section~\ref{3c.sec}. 

\subsubsection*{Normal-reference scale-invariant tests}\label{nrsit2c.sec}

As mentioned earlier, the Hotelling $T^2$-test is affine invariant under a linear transformation. 
However, the tests introduced by \cite{bai1996effect,Chen_2010,zhang2020simple,Zhang_2021,zhang2022further,zhang2022revisit} exhibit a lack of affine invariance. To address this issue, one approach involves scaling each variable by its sample standard deviation. This is equivalent to considering $(\overline{\bm{y}}_1-\overline{\bm{y}}_2)^{\top} \hat{\bm{D}}^{-1}(\overline{\bm{y}}_1-\overline{\bm{y}}_2)$ instead of $\|{\bm{y}}_1-\overline{\bm{y}}_2\|^2$ in the construction of a test statistic, where $\hat{\bm{D}}=\operatorname{diag}(\hat{\bm{\Sigma}})$ represents a diagonal matrix formed by the diagonal entries of the pooled sample covariance matrix $\hat{\bm{\Sigma}}$ as defined earlier. Such tests are referred to as scale-invariant tests, remaining unchanged under any scale transformation of high-dimensional data. In contrast, non-scale-invariant tests lack this property. Scale-invariant tests generally possess higher power, as they account for the diagonal variations of sample covariance matrices, making them preferred, albeit often requiring more stringent conditions.

\cite{Srivastava_2008} first proposed such a test statistic defined as:
\begin{equation}\label{T_SD}
T_{\mathrm{SD}}=\frac{n^{-1}n_1 n_2(\overline{\bm{y}}_1-\overline{\bm{y}}_2)^{\top} \hat{\bm{D}}^{-1}(\overline{\bm{y}}_1-\overline{\bm{y}}_2)-(n-4)^{-1}(n-2)p}{\sqrt{2[\mathrm{tr} (\hat{\bm{R}}^2)-p^2/(n-2)] c_{p, n}}},
\end{equation}
where $c_{p, n}$ is an adjustment coefficient such that $c_{p, n} \rightarrow 1$ in probability as $(n, p) \rightarrow \infty$. The authors suggested using $c_{p, n}=1+p^{-3 / 2} \operatorname{tr}(\hat{\bm{R}}^2)$, where $\hat{\bm{R}}=\hat{\bm{D}}^{-\frac{1}{2}} \hat{\bm{\Sigma}} \hat{\bm{D}}^{-\frac{1}{2}}$ is the sample correlation matrix. They demonstrated  that under TA1, $T_{\mathrm{SD}}$ converges to a standard normal distribution $N(0,1)$.  Additionally, $c_{p, n}$ is employed to enhance the convergence of $T_{\mathrm{SD}}$ to $N(0,1)$.

\cite{Srivastava_2013} extended the work of \cite{Srivastava_2008} to the two-sample BF problem for high-dimensional data and proposed the following test statistic:
\begin{equation}\label{T_SKK}
T_{\mathrm{SKK}}=\frac{n^{-1}n_1 n_2 \left(\overline{\bm{y}}_1-\overline{\bm{y}}_2\right)^{\top} \hat{\bm{D}}_n^{-1}\left(\overline{\bm{y}}_1-\overline{\bm{y}}_2\right)-p}{\sqrt{\hat{\sigma}^2 c_{n, p}^*}},
\end{equation}
where  $c_{n, p}^*=1+p^{-3 / 2}\operatorname{tr}(\hat{\bm{R}}_n^2)$ and 
$\hat{\sigma}^2=2\{\operatorname{tr}(\hat{\bm{R}}_n^2)-[(n_1-1)^{-1} n_2^2 \operatorname{tr}^2(\hat{\bm{D}}_n^{-1} \hat{\bm{\Sigma}}_1)+(n_2-1)^{-1} n_1^2 \operatorname{tr}^2(\hat{\bm{D}}_n^{-1} \hat{\bm{\Sigma}}_2)] / n^2\}$. Similar to \cite{Srivastava_2008}, $c_{n, p}^*$ is also an adjustment coefficient used to improve the convergence of $T_{\mathrm{SKK}}$ to $N(0,1)$. However, the asymptotic normality of $T_{\mathrm{SKK}}$ holds true only under the fulfillment of certain stringent conditions. In cases where these assumptions are not met, the corresponding normal approximation becomes inappropriate.

To overcome this problem,  \cite{Zhang_2020} proposed a new scale-invariant test statistic when $\bm{\Sigma}_1 = \bm{\Sigma}_2$, 
and 
 \cite{zhang2023two} constructed a normal-reference scale-invariant test for the two-sample  high-dimensional BF problem as follows:
\begin{equation}\label{T_ZZZ_SCALE2023}
T_{\mathrm{ZZZ}}=\frac{n_1 n_2}{n p}\left(\overline{\bm{y}}_1-\overline{\bm{y}}_2\right)^{\top} \hat{\bm{D}}_n^{-1}\left(\overline{\bm{y}}_1-\overline{\bm{y}}_2\right).
\end{equation}
There is a close connection between $T_{\mathrm{ZZZ}}$ \eqref{T_ZZZ_SCALE2023} and $T_{\mathrm{SKK}}$ \eqref{T_SKK} as seen from the expressions and for large samples, the distributions of $T_{\mathrm{ZZZ}}$ and $T_{\mathrm{SKK}}$ have similar shapes, either symmetric or skewed.  \cite{zhang2023two} showed that under TA1 or TA3, TA4, the null hypothesis, and $\log(p)=o(n_{\min})$, where $n_{\min}=\min(n_1,n_2),$  $\calL[T_{\mathrm{ZZZ}}(\calY_1,\calY_2)]=\calL[T_{\mathrm{ZZZ}}(\calY_1^*,\calY_2^*)]$ asymptotically and $\calL[T_{\mathrm{ZZZ}}(\calY_1^*,\calY_2^*)]=\calL(T_{\mathrm{ZZZ}}^*)$ with
    $T_{\mathrm{ZZZ}}^* = p^{-1} \sum_{r=1}^p \gamma_{n,p, r} A_r$, where $A_r,r=1,\ldots,p \stackrel{\text { i.i.d. }}{\sim} \chi_1^2$ and $\gamma_{n,p,r}$'s are defined in TA3.  Hence it is justified to use $\calL(T_{\mathrm{ZZZ}}^*)$ to approximate $\calL[T_{\mathrm{ZZZ}}(\calY_1,\calY_2)]$. Since the coefficients of $T_{\mathrm{ZZZ}}^*$ are all non-negative, they employed the W--S $\chi^2$-approximation as described in Section~\ref{2c.sec} to approximate $\calL(T_{\mathrm{ZZZ}}^*)$. Note that since $\operatorname{E}(T_{\mathrm{ZZZ}}^*)=1$,  $\calL(T_{\mathrm{ZZZ}}^*)$ can  actually be  approximated by $\calL(G)$ with $G\stackrel{d}{=} \chi_d^2/d$ where $d$ is the approximate degrees of freedom of the W--S $\chi^2$-approximation. The readers are referred to \cite{Zhang_2020} and \cite{zhang2023two} for more details.

\subsection{Tests for the GLHT problem}\label{GLHTproblem}

\subsubsection*{GLHT problem under high-dimensional MANOVA}\label{manova.sec}

 Testing whether the $k$ (when  $k>2$) mean vectors in \eqref{hypothesis} are the same  is also known as a one-way MANOVA testing problem, which can be treated as a special case of the following general linear hypothesis testing (GLHT) problem for high-dimensional data: 
\begin{equation}\label{GLHT}
H_0: \bm{G} \bm{M}=\bm{0}, \quad \text { vs. } \quad H_1: \bm{G} \bm{M} \neq \bm{0},
\end{equation}
where $\bm{M}=\left(\bm{\mu}_1, \ldots, \bm{\mu}_k\right)^{\top}$ is a $k \times p$ matrix collecting all the $k$ mean vectors and $\bm{G}: q \times k$ is a known full-rank coefficient matrix with $\operatorname{rank}(\bm{G})=q<k$. The GLHT problem \eqref{GLHT} can reduce to the one-way MANOVA problem when we set $\bm{G}$ to be either $\bm{G}_1=(\bm{I}_{k-1},-\bm{1}_{k-1})$  or  $\bm{G}_2=\left(-\bm{1}_{k-1}, \bm{I}_{k-1}\right)$, where  $\bm{1}_r$ denotes the $r$-dimensional vector of 1's. Actually, the GLHT problem \eqref{GLHT} is very general. When the null hypothesis in \eqref{hypothesis} is rejected, it is often of interest to further test if $\bm{\mu}_1=c_1 \bm{\mu}_3$ or if a contrast is zero, e.g., $c_2 \bm{\mu}_1-\left(c_2+c_3\right) \bm{\mu}_2+c_3 \bm{\mu}_3=0$ where $c_1, c_2$, and $c_3$ are some known constants. To write the above two testing problems in the form of \eqref{GLHT}, we just need to set $\bm{G}=\left(\bm{e}_{1, k}-c_1 \bm{e}_{3, k}\right)^{\top}$ and $\bm{G}=\left(c_2 \bm{e}_{1, k}-\left(c_2+c_3\right) \bm{e}_{2, k}+c_3 \bm{e}_{3, k}\right)^{\top}$ respectively where and throughout $\bm{e}_{r, l}$ denotes a unit vector of length $l$ with the $r$-th entry being 1 and others 0.

While the GLHT problem \eqref{GLHT} is of paramount importance, as emphasized in Section~\ref{sec:intro}, there exists a scarcity of dedicated articles addressing its testing aspects. Furthermore, the existing literature often imposes stringent assumptions on the underlying covariance matrices of the $k$ samples. It is noteworthy that the GLHT problem \eqref{GLHT} remains invariant under the following transformation of the coefficient matrix $\bm{G}$: $\bm{G} \rightarrow \bm{P} \bm{G}$, where $\bm{P}$ is any non-singular $q \times q$ matrix. In other words, non-singular transformations of the coefficient matrix $\bm{G}$ define the same hypothesis. Therefore, it is reasonable to expect that the proposed test should exhibit invariance under any non-singular transformations of $\bm{G}$. To achieve this, we will reformulate the GLHT problem \eqref{GLHT} into the following equivalent form:
\begin{equation}\label{GLHT_NEW2}
H_0: \bm{C} \bm{\mu}=\bm{0}, \quad \text { vs. } \quad H_1: \bm{C} \bm{\mu} \neq \bm{0},
\end{equation}
where $\bm{C}=[(\bm{G} \bm{B} \bm{G}^{\top})^{-1 / 2} \bm{G}]\otimes \bm{I}_p$, $\bm{B}=\operatorname{diag}(1 / n_1, \ldots, 1 / n_k)$, and $\bm{\mu}=(\bm{\mu}_1^{\top}, \ldots, \bm{\mu}_k^{\top})^{\top}$ with $\otimes$ denoting the Kronecker product operator. Let 
\begin{equation}\label{Omegank.equ}
    \bOmega_n=\COV(\bm{C} \hat{\bm{\mu}})=\bm{C}\bOmega\bm{C}^\top,
\end{equation}
where $\bOmega=\COV(\hat{\bm{\mu}})=\operatorname{diag}(\bm{\Sigma}_1/n_1, \ldots, \bm{\Sigma}_k/n_k):(k p) \times(k p)$ with $\hat{\bm{\mu}}=(\overline{\bm{y}}_1^{\top}, \ldots, \overline{\bm{y}}_k^{\top})^{\top}$ being the usual unbiased estimator of $\bm{\mu}$. For saving spaces, the detailed technical assumptions for the GLHT problem, which  can be regarded as generalizations of TA to more comprehensive problem settings, will be omitted  here. 

\cite{Zhang_2017} firstly constructed an $L^2$-norm-based test statistic for testing \eqref{GLHT_NEW2}, assuming  $\bm{\Sigma}_1=\cdots=\bm{\Sigma}_k$ and  \cite{zhang2022linear} extended the work for the heteroscedastic one-way MANOVA problem with the following test statistic: 
\begin{equation}\label{T_ZGZ}
T_{\mathrm{ZZG}}=\|\bm{C} \hat{\bm{\mu}}\|^2.
\end{equation}
They showed that under $H_0$ and some regularity conditions, we have $\calL[T_{\mathrm{ZZG}}(\calY_1,\ldots,\calY_k)]=\calL[T_{\mathrm{ZZG}}(\calY_1^*,\ldots,\calY_k^*)]$ asymptotically, and $\calL[T_{\mathrm{ZZG}}(\calY_1^*,\ldots,\calY_k^*)]$
is the same as the distribution of the $\chi^2$-type mixture $T_{\mathrm{ZZG}}^*=\sum_{r=1}^{qp} \lambda^*_{n, p, r} A_r, \; A_r\iidsim \chi_{1}^2$ where $\lambda^*_{n,p,r}$'s are the eigenvalues of $\bOmega_n$. Hence it is justified to use $\calL(T_{\mathrm{ZZG}}^*)$ to approximate $\calL[T_{\mathrm{ZZG}}(\calY_1,\ldots,\calY_k)]$, and $\calL(T_{\mathrm{ZZG}}^*)$ can be further approximated by using the W--S $\chi^2$-approximation as described in Section~\ref{2c.sec}.

To enhance  the test's performance in terms of size control, \cite{zhu2022linear} proposed a centralized $L^2$-norm-based test,  assuming $\bm{\Sigma}_1=\cdots=\bm{\Sigma}_k$ and  \cite{Zhang_2022heteroscedastic} extended the work for the heteroscedastic one-way MANOVA problem using the following statistic:
\begin{equation}\label{T_ZZ_3CBF}
T_{\mathrm{Z}^2}=\|\bm{C \hat{\mu}}\|^2-\sum_{i=1}^kh_{ii}\operatorname{tr}(\hat{\bm{\Sigma}}_i)/n_i,
\end{equation}
where $h_{ii}$ is the $i$-th diagonal entry of the matrix  $\bm{H}=\bm{G}^{\top}(\bm{G}\bm{B} \bm{G}^{\top})^{-1}\bm{G}$. \cite{Zhang_2022heteroscedastic} showed that, under $H_0$ and some regularity conditions, $\calL[T_{\mathrm{Z}^2}(\calY_1,\ldots,\calY_k)]=\calL[T_{\mathrm{Z}^2}(\calY_1^*,\ldots,\calY_k^*)]$ asymptotically, and $\calL[T_{\mathrm{Z}^2}(\calY_1^*,\ldots,\calY_k^*)]$
is the same as the distribution of the $\chi^2$-type mixture $T_{\mathrm{Z}^2}^*=\sum_{r=1}^{q p} \lambda^*_{n, p, r} A_r-\sum_{i=1}^k \sum_{r=1}^p [n_i(n_i-1)]^{-1}h_{ii} \lambda_{ir} B_{ir}$, 
where $\lambda^*_{n,p,r}$'s are the eigenvalues of $\bOmega_n$,   $A_r, r=1, \ldots,(q p) \stackrel{\text { i.i.d. }}{\sim}\chi_1^2$ and $B_{ir}, r=1, \ldots, p \stackrel{\text { i.i.d. }}{\sim} \chi_{n_i-1}^2$ are independent from each other. Since the unknown coefficients of  $T_{\mathrm{Z}^2}^*$ can be either positive or negative, \cite{Zhang_2022heteroscedastic} utilized  the 3-c matched $\chi^2$-approximation to approximate $\calL(T_{\mathrm{Z}^2}^*)$ as described in Section~\ref{3c.sec}. 

\subsubsection*{GLHT problem under high-dimensional linear regression}\label{reg.sec}

Several tests have been proposed for the high-dimensional linear regression model, expressed as $\mathbf{Y}=\mathbf{X}\bm{\Theta}+\bm{\epsilon}$. Here, $\mathbf{Y}=\left(\by_1, \ldots, \by_n\right)^{\top}$ is an $n \times p$ response matrix obtained by independently observing a $p$-dimensional response variable for $n$ subjects. The design matrix 
$\mathbf{X}$ is a known $n \times f$ full-rank matrix with $\operatorname{rank}(\mathbf{X})=f<n-2$. The parameter matrix $\boldsymbol{\Theta}$ is $f \times p$ and unknown, while the error matrix $\boldsymbol{\epsilon}=\left(\boldsymbol{\epsilon}_1, \ldots, \boldsymbol{\epsilon}_n\right)^{\top}$ is $n \times p$. Here, $\boldsymbol{\epsilon}_i$, for $i=1, \ldots, n$, are i.i.d. with a mean vector $\E(\boldsymbol{\epsilon}_i)=\mathbf{0}$ and a covariance matrix $\COV(\boldsymbol{\epsilon}_i)=\boldsymbol{\Sigma}$. Of interest is to test the following GLHT problem:
\begin{equation}\label{GLHT_LR}
H_0: \bm{\calC} \boldsymbol{\Theta}=\mathbf{0}, \quad \text { vs. } \quad H_1: \bm{\calC} \bm{\Theta} \neq \mathbf{0}, 
\end{equation}
where $ \bm{\calC} $ is a known matrix of size $q \times f$, with $\operatorname{rank}(\bm{\calC})=q<f$. It is worth to note that the above GLHT problem also includes one-way MANOVA or two-way MANOVA problems as special cases. For example, by setting $\boldsymbol{\Theta}=(\bm{\mu}_1,\ldots,\bm{\mu}_k)^\top$ and $\bm{\calC} = (\bm{I}_{k-1},-\bm{1}_{k-1})$, the GLHT problem (\ref{GLHT_LR}) reduces to the one-way MANOVA problem (\ref{hypothesis}). Note that the usual least squares estimator of $\boldsymbol{\Theta}$ is given by $\hat{\boldsymbol{\Theta}}=(\mathbf{X}^{\top} \mathbf{X})^{-1} \mathbf{X}^{\top} \mathbf{Y}$. Then, the variation matrices due to the hypothesis and error, denoted as $\mathbf{S}_h$ and $\mathbf{S}_e$, respectively, can be expressed as
$\mathbf{S}_h=(\bm{\calC} \hat{\mathbf{\Theta}})^{\top}[\bm{\calC}(\mathbf{X}^{\top} \mathbf{X})^{-1} \bm{\calC}^{\top}]^{-1} (\bm{\calC} \hat{\mathbf{\Theta}})=\mathbf{Y}^{\top} \mathbf{H}_X \mathbf{Y}$, and $\mathbf{S}_e=(\mathbf{Y}-\mathbf{X} \hat{\mathbf{\Theta}})^{\top}(\mathbf{Y}-\mathbf{X} \hat{\mathbf{\Theta}})=\mathbf{Y}^{\top}(\mathbf{I}_n-\mathbf{P}_X) \mathbf{Y}$,
where $\mathbf{P}_X=\mathbf{X}(\mathbf{X}^{\top} \mathbf{X})^{-1} \mathbf{X}^{\top}$ and $\mathbf{H}_X=\mathbf{X}(\mathbf{X}^{\top} \mathbf{X})^{-1} \bm{\calC}^{\top}[\bm{\calC}(\mathbf{X}^{\top} \mathbf{X})^{-1} \bm{\calC}^{\top}]^{-1} \bm{\calC}(\mathbf{X}^{\top} \mathbf{X})^{-1} \mathbf{X}^{\top}$ are two useful idempotent matrices of ranks $f$ and $q$, respectively.

Several tests have been proposed for high-dimensional Gaussian data, as evidenced by tests put forth by 
\cite{Fujikoshi_2004,Srivastava_2006,Schott_2007,yamada2012test}. These studies demonstrate that under the null hypothesis and certain regularity conditions, their respective test statistics exhibit asymptotic normal distribution.
It is noteworthy that the tests proposed by \cite{Fujikoshi_2004, Srivastava_2006, Schott_2007} lack scale invariance, potentially leading to reduced power, especially in high-dimensional data scenarios where the $p$ variables exhibit varying scales. This limitation underscores the necessity for scale-invariant tests. The subsequent introduction of scale-invariant tests, such as $T_{\mathrm{YS}}$ proposed by \cite{yamada2012test}, addresses this concern.

The insight provided by \cite{Zhu_2023} underscores the critical need to meticulously evaluate the suitability of employing normal approximation for test statistics. When the underlying distributional assumptions or conditions necessary for normal approximation are not satisfied, depending on it could yield inaccurate and potentially misleading test outcomes. Therefore, \cite{Zhu_2023} proposed a normal-reference scale-invariant test, which does not presuppose that the data adhere to a Gaussian distribution.  Their test statistic is given by: 
\begin{equation}\label{T_ZZZ_GLHT}
T_{\mathrm{Z}^3}=\frac{(n-f-2)}{(n-f)pq}\operatorname{tr}(\bm{S}_h\hat{\mathbf{D}}_e^{-1}),
\end{equation}
where $\hat{\mathbf{D}}_e=(n-f)^{-1}\operatorname{diag}(\mathbf{S}_e)$ and $\mathbf{R}_e=(n-f)^{-1}\hat{\mathbf{D}}_e^{-1/2}\mathbf{S}_e\hat{\mathbf{D}}_e^{-1/2}$. Let $\mathcal{Y}$ denote the response sample $\by_1,\ldots,\by_n$, and let $\mathcal{Y}^*$ denote $\mathcal{Y}$ when the sample $\by_1,\ldots,\by_n$ is treated as if it were normally distributed.
\cite{Zhu_2023} showed that under the null hypothesis and some regularity conditions, $\calL[T_{\mathrm{Z}^3}(\calY)]=\calL[T_{\mathrm{Z}^3}(\calY^*)]$ asymptotically, and $\calL[T_{\mathrm{Z}^3}(\calY^*)]$ is the same as the distribution of the central $\chi^2$-type mixture $T_{\mathrm{Z}^3}^* \stackrel{d}{=}(p q)^{-1} \sum_{r=1}^p \gamma_{p, r} A_r,  A_r \stackrel{i . i . d .}{\sim} \chi_q^2$, where $\gamma_{p, r}$'s are the eigenvalues of $\mathbf{R}=\mathbf{D}^{-1/2}\bSigma\mathbf{D}^{-1/2}$ in descending order with $\mathbf{D}=\operatorname{diag}(\bSigma)$.  Hence it is justified to use $\calL(T_{\mathrm{Z}^3}^*)$ to approximate $\calL[T_{\mathrm{Z}^3}(\calY)]$. Since $\E(T_{\mathrm{Z}^3}^*)=1$, $\calL(T_{\mathrm{Z}^3}^*)$ can actually be approximated by $\calL(G)$ with $G\stackrel{d}{=} \chi_d^2/d$  via the W--S $\chi^2$-approximation as described in Section~\ref{2c.sec}. The readers are referred to \cite{Zhu_2023} for more details.

\section{The HDNRA package in R}\label{sec:3}

 \pkg{HDNRA} offers optimized test statistics implemented in \proglang{C++} for both the two-sample problem, and the GLHT problem. For convenience, it includes two real datasets: \code{COVID19} (\citealt{thair2021transcriptomic}) and \code{corneal} (\citealt{smaga2019linear}). Additionally, it provides a suite of $21$ tests tailored for high-dimensional location testing, as outlined in Table~\ref{tab:overview}. These tests encompass seven normal-reference tests for the two-sample problem, five normal-reference tests for the GLHT problem, four normal-approximation-based tests (NABTs) for the two-sample problem, and five NABTs for the GLHT problem.

\begin{sidewaystable}[htbp]
\centering
\begin{tabular}{ l l l l l l }\toprule
Problem & \multicolumn{2}{c}{Approach}  &Function name & Test statistic & Reference\\
\midrule
\multirow{11}{*}{\shortstack[l] {Two-sample\\ problem}} & \multirow{7}{*}{\shortstack[l]{NRTs}} & \multirow{5}{*}{\shortstack[l]{2-c matched\\ $\chi^2$-approx.}}  & \code{ZGZC2020.TS.2cNRT()} & $T_{\mathrm{ZGZC}}$ \eqref{T_ZGZC} & \cite{zhang2020simple}\\
\cmidrule{4-6} & & & \code{ZZZ2020.TS.2cNRT()} & $T_{\mathrm{ZZZ}}$ \eqref{T_ZZZ_SCALE2023} & \cite{Zhang_2020}\\ 
\cmidrule{4-6} & & & \code{ZZGZ2021.TSBF.2cNRT()} & $T_{\mathrm{ZGZC}}$ \eqref{T_ZGZC} & \cite{Zhang_2021}\\
\cmidrule{4-6} & & & \code{ZWZ2023.TSBF.2cNRT()} & $T_{\mathrm{ZWZ}}$ \eqref{T_ZWZ} & \cite{zhu2022two}\\
\cmidrule{4-6} & & & \code{ZZZ2023.TSBF.2cNRT()} & $T_{\mathrm{ZZZ}}$ \eqref{T_ZZZ_SCALE2023} & \cite{zhang2023two}\\
\cmidrule{3-6} & & \multirow{2}{*}{\shortstack[l]{3-c matched\\ $\chi^2$-approx.}} & \code{ZZ2022.TS.3cNRT()} & $T_{\mathrm{BS}}$ \eqref{T_BS} & \cite{zhang2022revisit}\\
\cmidrule{4-6} & & & \code{ZZ2022.TSBF.3cNRT()} & $T_{\mathrm{ZZ}}$ \eqref{T_ZZ_CQ} & \cite{zhang2022further}\\
\cmidrule{2-6} & \multirow{4}{*}{\shortstack[l]{NABTs}} & & \code{BS1996.TS.NABT()} & $T_{\mathrm{BS}}$ \eqref{T_BS} & \cite{bai1996effect}\\
\cmidrule{4-6} & & & \code{SD2008.TS.NABT()} & $T_{\mathrm{SD}}$ \eqref{T_SD} & \cite{Srivastava_2008}\\
\cmidrule{4-6} & & & \code{CQ2010.TSBF.NABT()} & $T_{\mathrm{CQ}}$ \eqref{T_CQ}& \cite{Chen_2010}\\
\cmidrule{4-6} & & & \code{SKK2013.TSBF.NABT()} & $T_{\mathrm{SKK}}$ \eqref{T_SKK}& \cite{Srivastava_2013}\\
\midrule\multirow{10}{*}{ \shortstack[l] {GLHT\\ problem}} & \multirow{5}{*}{\shortstack[l]{NRTs}} & \multirow{3}{*}{\shortstack[l]{2-c matched\\ $\chi^2$-approx.}} 
 & \code{ZGZ2017.GLHT.2cNRT()} & $T_{\mathrm{ZZG}}$ \eqref{T_ZGZ} & \cite{Zhang_2017}\\
\cmidrule{4-6} & & & \code{ZZZ2022.GLHT.2cNRT()} & $T_{\mathrm{Z}^3}$ \eqref{T_ZZZ_GLHT} & \cite{Zhu_2023}\\
\cmidrule{4-6} & & & \code{ZZG2022.GLHTBF.2cNRT()} & $T_{\mathrm{ZZG}}$ \eqref{T_ZGZ} & \cite{zhang2022linear}\\
\cmidrule{3-6} & & \multirow{2}{*}{\shortstack[l]{3-c matched\\ $\chi^2$-approx.}} & \code{ZZ2022.GLHTBF.3cNRT()} & $T_{\mathrm{Z}^2}$ \eqref{T_ZZ_3CBF} & \cite{Zhang_2022heteroscedastic}\\
\cmidrule{4-6} & & & \code{ZZ2022.GLHT.3cNRT()} & $T_{\mathrm{Z}^2}$ \eqref{T_ZZ_3CBF} & \cite{zhu2022linear}\\
\cmidrule{2-6} &  \multirow{5}{*}{\shortstack[l]{NABTs}} & & \code{FHW2004.GLHT.NABT()} & $T_{\mathrm{FHW}}$ & \cite{Fujikoshi_2004}\\
\cmidrule{4-6} & & & \code{SF2006.GLHT.NABT()} & $T_{\mathrm{SF}}$  & \cite{Srivastava_2006}\\
\cmidrule{4-6} & & & \code{S2007.ks.NABT()} & $T_{\mathrm{S}}$ & \cite{Schott_2007}\\
\cmidrule{4-6} & & & \code{YS2012.GLHT.NABT()} & $T_{\mathrm{YS}} $  & \cite{yamada2012test}\\
\cmidrule{4-6} & & & \code{ZGZ2017.GLHTBF.NABT()} & $T_{\mathrm{Z}^2} \eqref{T_ZZ_3CBF}$ & \cite{Zhou_2017}\\
\bottomrule
\end{tabular}
\caption{\label{tab:overview} Overview of the tests available in \pkg{HDNRA}. Their details can be found  at \url{https://nie23wp8738.github.io/HDNRA/reference/index.html}.}
\end{sidewaystable}

\subsection{Dependencies}

\pkg{HDNRA} is an open-source software distributed under the GNU GPL-3 license. The core functionality of \pkg{HDNRA} is implemented in \proglang{C++} (\citealt{stroustrup2013c++}), using only standard libraries. In its implementation, extensive utilization of \proglang{C++11} features has been made. Apart from \pkg{Rcpp} (\citealt{eddelbuettel2011rcpp}), \pkg{RcppArmadillo} (\citealt{RcppArmadillo}), and \pkg{OpenMP} (\citealt{openmp2023}) for parallelization, the package has no strict dependencies. This ensures that the toolbox remains stable, self-contained, and efficient for reuse, with OpenMP significantly enhancing performance when dealing with large datasets.

\subsection{Installation}

The stable release version of \pkg{HDNRA} (\citealt{HDNRA}) is available on the Comprehensive R Archive Network (CRAN) at \url{https://cran.rstudio.com/web/packages/HDNRA/index.html}, and it can be installed using the following command:
\begin{CodeChunk}
\begin{CodeInput}
R> install.packages("HDNRA")
\end{CodeInput}
\end{CodeChunk}
Furthermore, the latest (development) version is accessible on GitHub at \url{https://github.com/nie23wp8738/HDNRA}, and installation can be performed using:
\begin{CodeChunk}
\begin{CodeInput}
R> install.packages("devtools")
R> devtools::install_github("nie23wp8738/HDNRA")
\end{CodeInput}
\end{CodeChunk}
Or
\begin{CodeChunk}
\begin{CodeInput}
R> install.packages("remotes")
R> remotes::install_github("nie23wp8738/HDNRA")
\end{CodeInput}
\end{CodeChunk}

The latest version $1.0.0$ of  \pkg{HDNRA}  from Github and version $4.3.1$ of \proglang{R} were used throughout this paper.
\subsection{Print}\label{print}
Due to the lack of an appropriate object to display our function results and in order to save display space, we have defined our own object of S3 class \proglang{NRtest}, which is motivated by \proglang{htest} in the package \pkg{EnvStats} (\citealt{EnvStats-book}), containing both required and optional components depending on the specifics of the hypothesis test, shown as follows:

\subsubsection*{Required Components:}
These components must be present in every \code{"NRtest"} object:
\begin{itemize}
    \item \code{statistic} The numeric scalar containing the value of the test statistic, with a \code{names} attribute indicating the name of the test statistic.
    \item \code{p.value} The numeric scalar containing the p-value for the test.
    \item \code{null.value} The character string indicating the null hypothesis.
    \item \code{alternative} The character string indicating the alternative hypothesis.
    \item \code{method} The character string giving the name of the test.
\end{itemize}

\subsubsection*{Optional Components:}
These components are included depending on the specifics of the hypothesis test performed:
\begin{itemize}
    \item \code{parameter} The numeric vector containing the estimated approximation parameter(s) associated with the approximation method. This vector has a \code{names} attribute describing its element(s).
    \item \code{sample.size} The numeric vector containing the number of observations in each group used for the hypothesis test.
    \item \code{sample.dimension} The numeric scalar containing the dimension of the dataset used for the hypothesis test.
    \item \code{estimation.method} The character string giving the name of approximation approach used to approximate the null distribution of the test statistic.
    \item \code{data.name} The character string describing the data set used in the hypothesis test.
\end{itemize}

\subsubsection*{Examples}
Example 1: Using \citealt{bai1996effect}'s test (two-sample problem)

\begin{CodeChunk}
\begin{CodeInput}
R>  NRtest.obj1 <- NRtest.object(
R>    statistic = c("T[BS]" = 2.208),
R>    p.value = 0.0136,
R>    method = "Bai and Saranadasa (1996)'s test",
R>    data.name = "group1 and group2",
R>    null.value = "Two mean vectors are equal",
R>    alternative = "Two mean vectors are not equal",
R>    parameter = NULL,
R>    estimate = NULL,
R>    sample.size = c(n1 = 24, n2 = 26),
R>    sample.dimension = 20460,
R>    estimation.method = "Normal approximation"
R>  )
R>  print(NRtest.obj1)
\end{CodeInput}
\begin{CodeOutput}
Results of Hypothesis Test
--------------------------

Test name:                       Bai and Saranadasa (1996)'s test

Null Hypothesis:                 Two mean vectors are equal

Alternative Hypothesis:          Two mean vectors are not equal

Data:                            group1 and group2

Sample Sizes:                    n1 = 24
                                 n2 = 26

Sample Dimension:                20460

Test Statistic:                  T[BS] = 2.208

Approximation method to the      Normal approximation
null distribution of T[BS]: 

P-value:                         0.0136
\end{CodeOutput}
\end{CodeChunk}

Example 2: Using \citealt{Fujikoshi_2004}'s test (GLHT problem)

\begin{CodeChunk}
\begin{CodeInput}
R>  NRtest.obj2 <- NRtest.object(
R>    statistic = c("T[FHW]" = 6.4015),
R>    p.value = 0,
R>    method = "Fujikoshi et al. (2004)'s test",
R>    data.name = "Y",
R>    null.value = "The general linear hypothesis is true",
R>    alternative = "The general linear hypothesis is not true",
R>    estimate = NULL,
R>    sample.size = c(n1 = 43, n2 = 14, n3 = 21, n4 = 72),
R>    sample.dimension = 2000,
R>    estimation.method = "Normal approximation"
R>  )
R>  print(NRtest.obj2)
\end{CodeInput}
\begin{CodeOutput}
Results of Hypothesis Test
--------------------------

Test name:                       Fujikoshi et al. (2004)'s test

Null Hypothesis:                 The general linear hypothesis is true

Alternative Hypothesis:          The general linear hypothesis is not true

Data:                            Y

Sample Sizes:                    n1 = 43
                                 n2 = 14
                                 n3 = 21
                                 n4 = 72

Sample Dimension:                2000

Test Statistic:                  T[FHW] = 6.4015

Approximation method to the      Normal approximation
null distribution of T[FHW]: 

P-value:                         0
\end{CodeOutput}
\end{CodeChunk}
\subsection{Usage and Contribution}

The \textbf{HDNRA} package is designed for academic research and real-world applications, serving as a toolbox for high-dimensional location testing in \proglang{R}. The package includes functions for established tests and provides two high-dimensional datasets. These functions generate $p$-values, test statistics, and approximate parameters, enabling efficient detection of differences in mean vectors between populations.

By examining the approximate degrees of freedom, users can determine whether the null distribution is normal or non-normal, aiding in assessing test reliability. The included datasets are also suitable for other high-dimensional analyses.

The \textbf{HDNRA} package can be used alongside other packages for high-dimensional analysis (HDA), as mean testing typically serves as an initial step in HDA.

All code is open source, and the development version is available on GitHub at \url{https://github.com/nie23wp8738/HDNRA}. Contributions are welcome via GitHub issues and pull requests.

\section{Practical implementation through examples}\label{sec:4}

\subsection{Data}\label{data}

Once \pkg{HDNRA} is installed and loaded, the datasets \proglang{COVID19} and \proglang{corneal} are available (via lazy data mechanism).

\subsubsection*{The COVID-19 data}

For illustrative purposes, the dataset pertaining to COVID-19 available on NCBI (\url{https://www.ncbi.nlm.nih.gov/geo/query/acc.cgi?acc=GSE152641}) with ID GSE152641 was utilized. As documented by  \cite{thair2021transcriptomic}, this data set profiled peripheral blood from 24 healthy controls and 62 prospectively enrolled patients with community-acquired lower respiratory tract infection by SARS-COV-2 within the first 24 hours of hospital admission using RNA sequencing. Each RNA sequencing transcriptome profile has $20460$ measurements. 
\begin{CodeChunk}
\begin{CodeInput}
R> library("HDNRA")
R> data("COVID19")
R> dim(COVID19)
\end{CodeInput}
\begin{CodeOutput}
[1]    87 20460
\end{CodeOutput}
%We can divide the original data into healthy control group $1$ and COVID-$19$ patients group $2$.  We get two matrices with orders $24 \times 20460$ and $62 \times 20460$.
\begin{CodeInput}
R> group1 <- as.matrix(COVID19[c(2:19, 82:87), ]) ## healthy group
R> dim(group1)
\end{CodeInput}
\begin{CodeOutput}
[1]    24 20460
\end{CodeOutput}
\begin{CodeInput}
R> group2 <- as.matrix(COVID19[-c(1:19, 82:87), ]) ## COVID-19 patients
R> dim(group2)
\end{CodeInput}
\begin{CodeOutput}
[1]    62 20460
\end{CodeOutput}
\end{CodeChunk}

\begin{figure}[!ht]
\centering
        \includegraphics[width=0.7\textwidth]{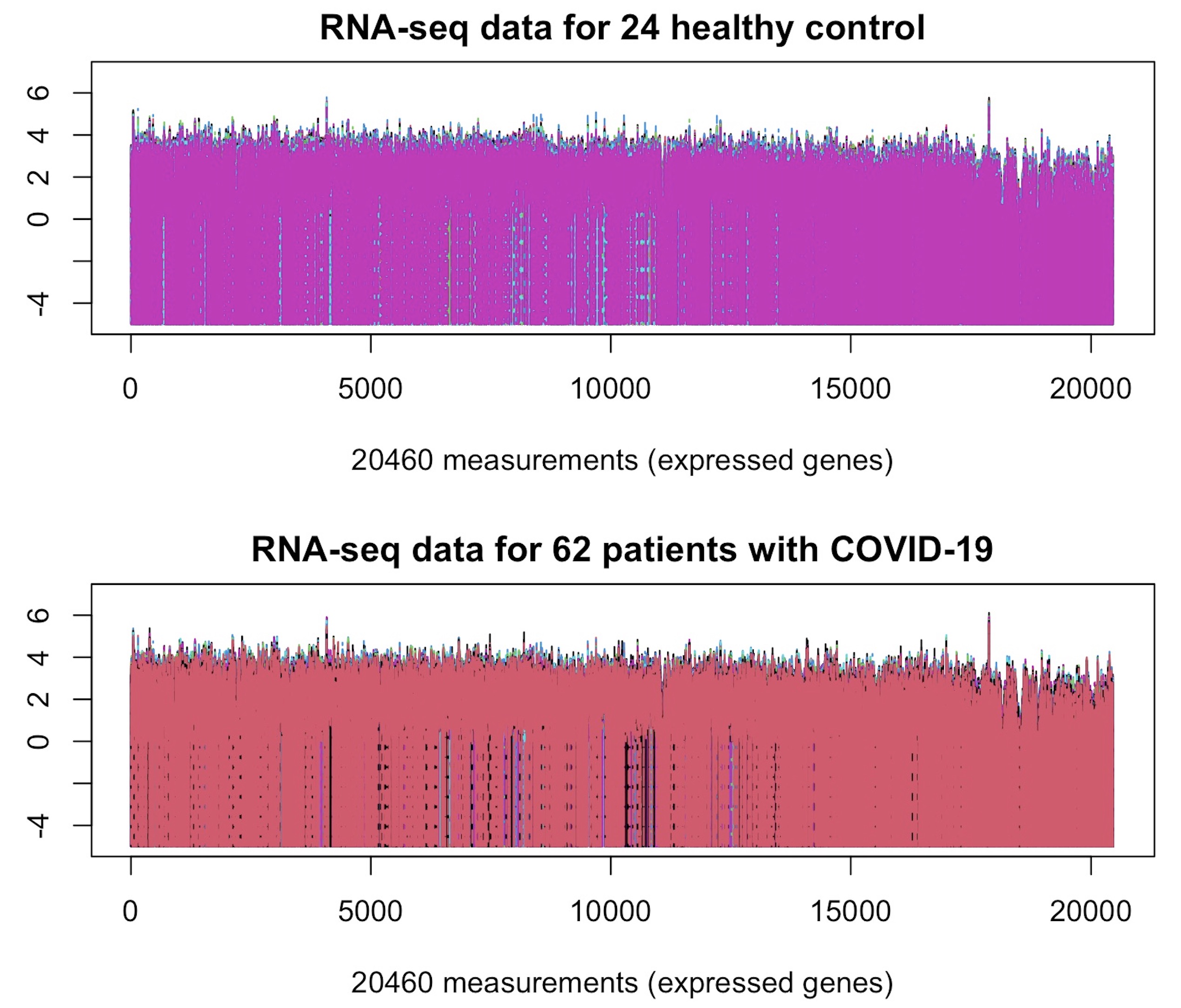}
        \caption{\label{fig:COVID19} RNA-seq data for patients with COVID-19 versus healthy control.}
\end{figure}

Taking into account the presence of null values in the dataset and  the maximum values in two groups, Figure \ref{fig:COVID19} displays the base $10$ logarithm of the RNA sequencing transcriptome profile measurements, for both the healthy control group and the COVID-19 patients. Distinguishing directly between these two groups based on the figure proves challenging.

\subsubsection*{The corneal data}

\begin{figure}[!ht]
\centering
        \includegraphics[width=0.7\textwidth]{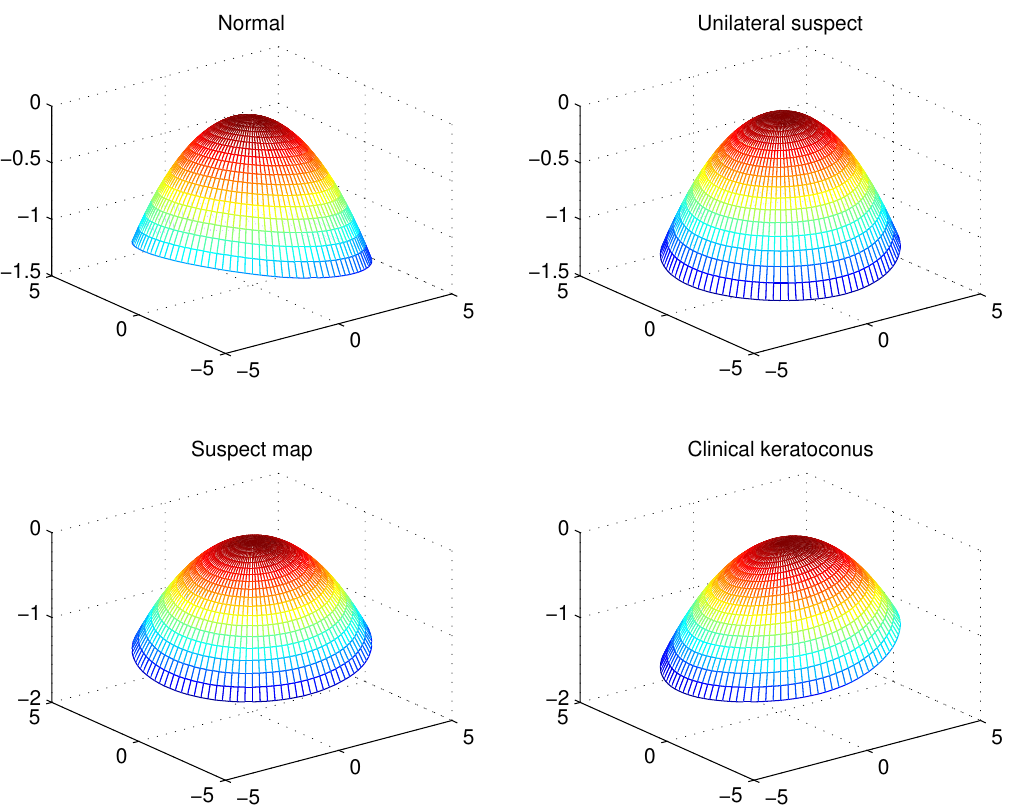}
        \caption{\label{fig:coreal} Examples of the corneal surfaces in the four cornea groups.}
\end{figure}

The \proglang{corneal} dataset from \cite{smaga2019linear} was acquired during a keratoconus study, which is a collaborative project involving Ms. Nancy Tripoli and Dr. Kenneth L. Cohen of Department of Ophthalmology at the University of North Carolina, Chapel Hill. After reconstructing the corneal surfaces, as detailed in \cite{smaga2019linear},  the \proglang{corneal} dataset comprises fitted feature vectors with a dimension of 2000 for each of the 150 corneal surfaces.
\begin{CodeChunk}
\begin{CodeInput}
R> data("corneal")
R> dim(corneal)
\end{CodeInput}
\begin{CodeOutput}
[1]  150 2000
\end{CodeOutput}
\end{CodeChunk}
In the \proglang{corneal} dataset, the observations from the normal group occupy the first 43 rows, followed by those from the unilateral suspect group in the next 14 rows. Subsequently, there are 21 rows containing observations from the suspect map group, with the observations from the clinical keratoconus group located in the last 72 rows. Figure \ref{fig:coreal} displays the examples of the corneal surfaces in the four cornea groups. Similarly, it is not feasible to directly distinguish between the four groups depicted in Figure \ref{fig:coreal}.
\begin{CodeChunk}
\begin{CodeInput}
R> group1 <- as.matrix(corneal[1:43, ]) ## normal group
R> dim(group1)
\end{CodeInput}
\begin{CodeOutput}
[1]   43 2000
\end{CodeOutput}
\end{CodeChunk}
\begin{CodeChunk}
\begin{CodeInput}
R> group2 <- as.matrix(corneal[44:57, ]) ## unilateral suspect group
R> dim(group2)
\end{CodeInput}
\begin{CodeOutput}
[1]   14 2000
\end{CodeOutput}
\end{CodeChunk}
\begin{CodeChunk}
\begin{CodeInput} 
R> group3 <- as.matrix(corneal[58:78, ]) ## suspect map group
R> dim(group3)
\end{CodeInput}
\begin{CodeOutput}
[1]   21 2000
\end{CodeOutput}
\end{CodeChunk}
\begin{CodeChunk}
\begin{CodeInput}
R> group4 <- as.matrix(corneal[79:150, ]) ## clinical keratoconus group
R> dim(group4)
\end{CodeInput}
\begin{CodeOutput}
[1]   72 2000
\end{CodeOutput}
\end{CodeChunk}

\subsection{Functions for the two-sample problem}\label{2sfun.sec}
In this section, we will illustrate the application of  \pkg{HDNRA}  for addressing the two-sample high-dimensional location testing problem using the \proglang{COVID19} dataset.\\ 

As detailed  in Table~\ref{tab:overview},  \pkg{HDNRA}  encompasses 11 testing procedures  for the two-sample problem. All the functions are controlled by the following two arguments:
\begin{itemize}
    \item \code{y1} The data matrix ($n_1$ by $p$) from the first population. Each row represents a $p$-dimensional observation.
    \item \code{y2}  The data matrix ($n_2$ by $p$) from the second population. Each row  represents a $p$-dimensional observation.
\end{itemize}
As described in Section~\ref{data}, it is of interest and worthwhile to check whether those prospectively enrolled patients with COVID-19 and healthy controls have the same mean RNA sequencing transcriptome profiles.
\begin{CodeChunk}
\begin{CodeInput}
R> data("COVID19")
R> group1 <- as.matrix(COVID19[c(2:19, 82:87), ])
R> group2 <- as.matrix(COVID19[-c(1:19, 82:87), ])
\end{CodeInput}
\end{CodeChunk}

The function returns an object of S3 class \proglang{NRtest} as described in Section~\ref{print}.
The \code{parameters} component varies across different methods.
For example, for the normal-reference tests with  2-c matched $\chi^2$-approximation, i.e., \code{ZGZC2020.TS.2cNRT()} and \code{ZZGZ2021.TSBF.2cNRT()},  we have
\begin{itemize}
    \item \code{df}  estimated approximate degrees of freedom of the test
    \item  \code{beta}  estimated parameters $\beta$ used in 2-c matched $\chi^2$-approximation
\end{itemize}
which are obtained from (\ref{2c}). 

The corresponding code for \cite{zhang2020simple}'s test is:
\begin{CodeChunk}
\begin{CodeInput}
R> ZGZC2020.TS.2cNRT(group1,group2)
\end{CodeInput}
\end{CodeChunk}
and we can get the following results in less than $0.2$ seconds:
\begin{CodeChunk}
\begin{CodeInput}
Results of Hypothesis Test
--------------------------

Test name:                       Zhang et al. (2020)'s test

Null Hypothesis:                 Difference between two mean vectors is 0

Alternative Hypothesis:          Difference between two mean vectors is not 0

Data:                            group1 and group2

Sample Sizes:                    n1 = 24
                                 n2 = 62

Sample Dimension:                20460

Test Statistic:                  T[ZGZC] = 228972526332

Approximation method to the      2-c matched chi^2-approximation
null distribution of T[ZGZC]: 

Approximation parameter(s):      df   = 2.605400e+00
                                 beta = 2.965057e+10

P-value:                         0.03771277
\end{CodeInput}
\end{CodeChunk}
 For the normal-reference scale-invariant tests, i.e., \code{ZZZ2020.TS.2cNRT()} and \code{ZZZ2023.TSBF.2cNRT()}, we can get
 \begin{itemize}
    \item \code{df} estimated approximate degrees of freedom of the test 
\end{itemize}
which has been described in Section~\ref{twosample}. 
The corresponding code and result using \code{ZZZ2020.TS.2cNRT()} are:
\begin{CodeChunk}
\begin{CodeInput}
R> ZZZ2020.TS.2cNRT(group1,group2) 
\end{CodeInput}
\begin{CodeOutput}
Results of Hypothesis Test
--------------------------

Test name:                       Zhang et al. (2020)'s test

Null Hypothesis:                 Difference between two mean vectors is 0

Alternative Hypothesis:          Difference between two mean vectors is not 0

Data:                            group1 and group2

Sample Sizes:                    n1 = 24
                                 n2 = 62

Sample Dimension:                20460

Test Statistic:                  T[ZZZ] = 5.2221

Approximation method to the      2-c matched chi^2-approximation
null distribution of T[ZZZ]: 

Approximation parameter(s):      df = 11.5033

P-value:                         1.416134e-08
\end{CodeOutput}
\end{CodeChunk}
For the normal-reference $F$-type test, i.e., \code{ZWZ2023.TSBF.2cNRT()}, we will obtain the following two parameters:
\begin{itemize}
    \item \code{df1} estimated approximate degrees of freedom $d_1$ 
\item \code{df2} estimated approximate degrees of freedom $d_2$ 
\end{itemize}
Below is the code that utilizes \cite{Zhu_2023}'s test:
\begin{CodeChunk}
\begin{CodeInput}
R> ZWZ2023.TSBF.2cNRT(group1,group2)
\end{CodeInput}
\begin{CodeOutput}
Results of Hypothesis Test
--------------------------

Test name:                       Zhu et al. (2023)'s test

Null Hypothesis:                 Difference between two mean vectors is 0

Alternative Hypothesis:          Difference between two mean vectors is not 0

Data:                            group1 and group2

Sample Sizes:                    n1 = 24
                                 n2 = 62

Sample Dimension:                20460

Test Statistic:                  T[ZWZ] = 4.1877

Approximation method to the      2-c matched chi^2-approximation
null distribution of T[ZWZ]: 

Approximation parameter(s):      df1 =   2.7324
                                 df2 = 171.7596

P-value:                         0.008672887
\end{CodeOutput}
\end{CodeChunk}

For the normal-reference tests with 3-c matched $\chi^2$-approximation, i.e., \code{ZZ2022.TS.3cNRT()}  and \code{ZZ2022.TSBF.3cNRT()}, we have
\begin{itemize}
    \item \code{df} estimated approximate degrees of freedom of the test 
    \item \code{beta0}  estimated parameter $\beta_0$ used in 3-c matched $\chi^2$-approximation
    \item \code{beta1}  estimated parameter $\beta_1$ used in 3-c matched $\chi^2$-approximation
\end{itemize}
which are obtained from (\ref{3c}).  The code for applying the test proposed by \cite{zhang2022revisit} is as follows:
\begin{CodeChunk}
\begin{CodeInput}
R> ZZ2022.TS.3cNRT(group1,group2)
\end{CodeInput}
\begin{CodeOutput}
Results of Hypothesis Test
--------------------------

Test name:                       Zhang and Zhu (2022)'s test

Null Hypothesis:                 Difference between two mean vectors is 0

Alternative Hypothesis:          Difference between two mean vectors is not 0

Data:                            group1 and group2

Sample Sizes:                    n1 = 24
                                 n2 = 62

Sample Dimension:                20460

Test Statistic:                  T_ZZ = 1.51016e+11

Approximation method to the      3-c matched chi^2-approximation
null distribution of T_ZZ: 

Approximation parameter(s):      df    =  1.731300e+00
                                 beta0 = -6.363520e+10
                                 beta1 =  3.675598e+10

P-value:                         0.04105057
\end{CodeOutput}
\end{CodeChunk}
We just follow the same pattern and will get the result of \code{ZZ2022.TSBF.3cNRT()} very quickly.

For the other existing tests, such as \code{SD2008.TS.NABT()} and \code{SKK2013.TSBF.NABT()}, we have
\begin{itemize}
    \item \code{cpn} calculated adjustment coefficient used in the test
\end{itemize}
The corresponding code for \cite{Srivastava_2013}'s test is:
\begin{CodeChunk}
\begin{CodeInput}
R> SKK2013.TSBF.NABT(group1,group2)
\end{CodeInput}
\begin{CodeOutput}
Results of Hypothesis Test
--------------------------

Test name:                       Srivastava et al. (2013)'s test

Null Hypothesis:                 Difference between two mean vectors is 0

Alternative Hypothesis:          Difference between two mean vectors is not 0

Data:                            group1 and group2

Sample Sizes:                    n1 = 24
                                 n2 = 62

Sample Dimension:                20460

Test Statistic:                  T[SKK] = 2.8966

Approximation method to the      Normal approximation
null distribution of T[SKK]: 

Approximation parameter(s):      Adjustment coefficient = 17.9488

P-value:                         0.001886357
\end{CodeOutput}
\end{CodeChunk}
Besides, implementing the test suggested by \cite{bai1996effect} can be achieved as follows:
\begin{CodeChunk}
\begin{CodeInput}
R> BS1996.TS.NABT(group1,group2)
\end{CodeInput}
\begin{CodeOutput}
 Results of Hypothesis Test
--------------------------

Test name:                       Bai and Saranadasa (1996)'s test

Null Hypothesis:                 Difference between two mean vectors is 0

Alternative Hypothesis:          Difference between two mean vectors is not 0

Data:                            group1 and group2

Sample Sizes:                    n1 = 24
                                 n2 = 62

Sample Dimension:                20460

Test Statistic:                  T[BS] = 2.208

Approximation method to the      Normal approximation
null distribution of T[BS]: 

P-value:                         0.01362284
\end{CodeOutput}
\end{CodeChunk}
The \proglang{R} code and test results of the remaining two-sample tests are given in Appendix~\ref{app:ts}.

To assess the performance of all 11 testing procedures applied to the \proglang{COVID19} dataset, the test results are presented in Table~\ref{tab:ts}. It is observed that all the tests reject the null hypothesis at a 5\% significance level, indicating significant differences in the mean transcriptome profiles between the two groups of the COVID-19 data. However, examining the "d.f." column reveals that all the estimated approximate degrees of freedom are small. This suggests that relying on the normal approximation to the null distribution, as done by \cite{bai1996effect, Srivastava_2008, Chen_2010, Srivastava_2013}, is generally not  adequate, and consequently, their $p$-values may not be reliable. It is also worthwhile to note that, for those normal-reference tests, the $p$-values obtained by \code{ZGZC2020.TS.2cNRT()} and \code{ZZ2022.TS.3cNRT()}, which show a similar magnitude, are notably larger compared to the $p$-values from \code{ZZGZ2021.TSBF.2cNRT()}, \code{ZWZ2023.TSBF.2cNRT()}, and \code{ZZ2022.TSBF.3cNRT()}. This discrepancy likely arises from the fact that the former two tests are based on the equal-covariance matrix assumption, while the latter three tests are not. In addition, those two normal-reference scale-invariant tests, namely, \code{ZZZ2020.TS.2cNRT()} and \code{ZZZ2023.TSBF.2cNRT()}, yield significantly smaller magnitudes than the aforementioned normal-reference non-scale-invariant tests. It is known that scale-invariant tests generally exhibit higher power than non-scale-invariant tests. However,  it is essential to note that the power gain of a scale-invariant test to a non-scale invariant test is not a free lunch.  Further details can be found in \cite{Zhang_2020,zhang2023two}.

\begin{table}[!ht]
\centering
\begin{tabular}{lllll}
\toprule
\multicolumn{2}{c}{Approach}  &Function name & $p$-value & d.f. \\
\midrule \multirow{7}{*}{\shortstack[l]{NRTs}} & \multirow{5}{*}{2-c matched $\chi^2$-approx.} & \code{ZGZC2020.TS.2cNRT()} & 0.0377 & 2.61 \\
\cmidrule{3-5} & & \code{ZZZ2020.TS.2cNRT()} & 
$<0.0001$ & 11.50 \\
\cmidrule{3-5} & & \code{ZZGZ2021.TSBF.2cNRT()} & 0.0069 & 2.73 \\
\cmidrule{3-5} & & \code{ZWZ2023.TSBF.2cNRT()} & 0.0087 &  2.73 \\
\cmidrule{3-5} & & \code{ZZZ2023.TSBF.2cNRT()} & 
$<0.0001$ & 10.13 \\
\cmidrule{2-5} & \multirow{2}{*}{3-c matched $\chi^2$-approx.} & \code{ZZ2022.TS.3cNRT()} & 0.0411 & 1.73 \\
\cmidrule{3-5} & & \code{ZZ2022.TSBF.3cNRT()} & 0.0092 & 1.91 \\
\midrule \multirow{4}{*}{\shortstack[l]{NABTs} } & & \code{BS1996.TS.NABT()} & 0.0136 & -- \\
\cmidrule{3-5} & & \code{SD2008.TS.NABT()} & 0.0051 & -- \\
\cmidrule{3-5} & & \code{CQ2010.TSBF.NABT()} & 0.0002 & -- \\
\cmidrule{3-5} & & \code{SKK2013.TSBF.NABT()} & 0.0019 & -- \\
\bottomrule
\end{tabular}
\caption{\label{tab:ts} Testing the equality of the mean transcriptome profiles of the two groups of the COVID-19 data  using  \pkg{HDNRA}.}
\end{table}

To further demonstrate the accuracy of the considered tests for high-dimensional two-sample problem, we use this \proglang{COVID19} data set to calculate the empirical sizes of these testing procedures. The empirical size is obtained from 10000 runs. In each run, we randomly split the 62 COIVD-19 patients into two groups of equal size. We calculate the empirical size as the proportion of times the $p$-value is smaller than the nominal level $\alpha = 5\%$ or $10\%$ based on the 10000 independent runs. The empirical sizes of the test procedures are presented in Table~\ref{tab:ts_size}, with the corresponding code for obtaining these empirical sizes provided in Appendix~\ref{app:ts_size}.

Several conclusions can be drawn from Table~\ref{tab:ts_size}. Firstly, in terms of size control, normal-reference tests generally outperform other existing tests. Tests by \cite{bai1996effect} and \cite{Chen_2010} demonstrate liberal behavior with empirical sizes around 6.5\% when the nominal level $\alpha=5\%$. Conversely, tests by \cite{Srivastava_2008} and \cite{Srivastava_2013} are notably conservative, with very small empirical sizes. This is unsurprising given the small estimated degrees of freedom and large adjustment coefficient values. Consequently, the normal approximation utilized by these four competitors is considered inappropriate. Secondly, among normal-reference tests, the two scale-invariant tests by \cite{Zhang_2020} and \cite{zhang2023two} exhibit larger empirical sizes compared to other normal-reference non-scale-invariant tests. As discussed previously, while scale-invariant tests may offer increased power compared to non-scale-invariant tests, this power gain often leads to larger empirical sizes. Thirdly, among normal-reference tests, the two tests employing 2-c $\chi^2$-approximation by \cite{zhang2020simple} and \cite{Zhang_2021} perform the best. This outcome was anticipated as more information is incorporated into these methods, as discussed in \cite{zhang2022further, zhang2022revisit,Zhu_2023}, leading to more accurate results. However, it is crucial to note that obtaining larger samples is essential for the application of these methods, and the sample size of 62 in this study may not be sufficient to fully leverage their potential.

\begin{table}[!ht]
\centering
\begin{tabular}{lllll}
\toprule & \multicolumn{2}{ c }{Empirical size (in \%)} & \multicolumn{2}{ c }{Parameters } \\
\cmidrule { 2 - 3} \cmidrule { 4 - 5}
& $\alpha = 5\%$ & $\alpha = 10\%$ & d.f. & cpn \\
\midrule \code{ZGZC2020.TS.2cNRT()} & 4.81 & 10.30 & 2.55 & -- \\
\midrule \code{ZZZ2020.TS.2cNRT()} & 6.17 & 10.39 & 12.00 & -- \\
\midrule \code{ZZGZ2021.TSBF.2cNRT()} & 4.89  & 10.40 & 2.55 & -- \\
\midrule \code{ZWZ2023.TSBF.2cNRT()} & 4.34 & 9.53 & 2.55 & -- \\
\midrule \code{ZZZ2023.TSBF.2cNRT()} & 6.17  & 10.39 & 12.02 & 15.42 \\
\midrule\code{ZZ2022.TS.3cNRT()} & 4.49  & 10.22 & 1.76 & -- \\
\midrule \code{ZZ2022.TSBF.3cNRT()} & 4.52 & 10.22 & 1.80 & -- \\
\midrule \code{BS1996.TS.NABT()} & 6.49 & 10.35 & -- & -- \\
\midrule \code{CQ2010.TSBF.NABT()} & 6.65  & 10.53 & -- & -- \\
\midrule \code{SD2008.TS.NABT()} & 0.16  & 0.55 & -- & 15.42 \\
\midrule \code{SKK2013.TSBF.NABT()} & 0.17  & 0.57 & -- & 15.42 \\
\bottomrule
\end{tabular}
\caption{\label{tab:ts_size} Comparison of the empirical sizes (in $\%$) of the two-sample testing procedures  using  \pkg{HDNRA}.}
\end{table}

\subsection{Functions for the GLHT problem}\label{glhtfun.sec}

 As mentioned in Section~\ref{GLHTproblem}, the GLHT problem (\ref{GLHT}) or (\ref{GLHT_LR}) is very general and also includes the well-known one-way MANOVA testing problem as a special case. In this section, we first illustrate the  application of \pkg{HDNRA} for addressing the one-way MANOVA problem with the \proglang{corneal} dataset, and then followed by some contrast tests. As introduced in Section~\ref{data}, there are four groups in the \proglang{corneal} dataset, i.e., the normal group, the unilateral suspect group, the suspect map group, and the clinical keratoconus group. Of interest is to check whether the keratoconus disease has a strong impact on the shapes of the corneal surfaces, i.e., whether the four corneal surface groups have the same mean corneal surface.
\begin{CodeChunk}
\begin{CodeInput}
R> data("corneal")
R> group1 <- as.matrix(corneal[1:43, ]) ## the normal group
R> group2 <- as.matrix(corneal[44:57, ]) ## the unilateral suspect group
R> group3 <- as.matrix(corneal[58:78, ]) ## the suspect map group
R> group4 <- as.matrix(corneal[79:150, ]) ## the clinical keratoconus group
\end{CodeInput}
\end{CodeChunk}
As detailed in Table~\ref{tab:overview}, the package includes 10 tests for the GLHT problem. In contrast to the inputs of functions for the two-sample problem, those for the GLHT problem lack neat and uniform organization. They can be categorized into two classes. The first class includes \code{ZGZ2017.GLHTBF.NABT()}, 
\code{ZZ2022.GLHTBF.3cNRT()}, \code{ZZG2022.GLHTBF.2cNRT()},

\code{ZGZ2017.GLHT.2cNRT()}, and \code{ZZ2022.GLHT.3cNRT()} which are governed by the following four arguments:

\begin{itemize}
    \item \code{Y}\;	a list of $k$ data matrices. The $i$th element represents the data matrix ($n_i\times p$) from the $i$th population with each row representing a $p$-dimensional observation.
  \item \code{G}\; a known full-rank coefficient matrix $(q\times k)$ with $\operatorname{rank}(\boldsymbol{G})< k$.
\item \code{n}\;	 a vector of $k$ sample sizes. The $i$th element represents the sample size of group $i$, $n_i$. 
\item \code{p}\; the dimension of data.
\end{itemize}

The following code combines the four groups from the \texttt{corneal} dataset into a list named \texttt{Y}. This list is then used to define other key parameters, such as the sample size vector \texttt{n}, the number of features \texttt{p}, and the contrast matrix \texttt{G} for hypothesis testing.
\begin{CodeChunk}
\begin{CodeInput}
R> p <- dim(corneal)[2]
R> k <- 4
R> Y <- list()
R> Y[[1]] <- group1
R> Y[[2]] <- group2
R> Y[[3]] <- group3
R> Y[[4]] <- group4
R> n <- c(nrow(Y[[1]]),nrow(Y[[2]]),nrow(Y[[3]]),nrow(Y[[4]]))
R> G <- cbind(diag(k-1),rep(-1,k-1))
\end{CodeInput}
\end{CodeChunk}

Each of the 10 functions designed for the GLHT problem produces a (list) object belonging to the S3 class \proglang{NRtest}. This object includes three elements: \code{p.value}, \code{statistic}, and \code{parameters}, as detailed in Section~\ref{2sfun.sec}. The interpretation of the \code{parameters} component may vary between different methods, but it can be understood in a manner similar to that described in Section~\ref{2sfun.sec}. As an illustration, consider \cite{zhang2022linear}'s test, which is a normal-reference test with a 2-c matched $\chi^2$-approximation. The corresponding code for utilizing \code{ZZG2022.GLHTBF.2cNRT()} is as follows:
\begin{CodeChunk}
\begin{CodeInput}
R> ZZG2022.GLHTBF.2cNRT(Y,G,n,p)
\end{CodeInput}
\begin{CodeOutput}
Results of Hypothesis Test
--------------------------

Test name:                       Zhang et al. (2022)'s test

Null Hypothesis:                 The general linear hypothesis is true

Alternative Hypothesis:          The general linear hypothesis is not true

Data:                            Y

Sample Sizes:                    n1 = 43
                                 n2 = 14
                                 n3 = 21
                                 n4 = 72

Sample Dimension:                2000

Test Statistic:                  T[ZZG] = 159.7325

Approximation method to the      2-c matched chi^2-approximation
null distribution of T[ZZG]: 

Approximation parameter(s):      df   = 6.1652
                                 beta = 6.1464

P-value:                         0.0002577084
\end{CodeOutput}
\end{CodeChunk}

To implement \cite{Zhang_2022heteroscedastic}'s test, which is a normal-reference test with a 3-c matched $\chi^2$-approximation, we can use the following code:
\begin{CodeChunk}
\begin{CodeInput}
R> ZZ2022.GLHTBF.3cNRT(Y,G,n,p)
\end{CodeInput}
\begin{CodeOutput}
Results of Hypothesis Test
--------------------------

Test name:                       Zhang and Zhu (2022)'s test

Null Hypothesis:                 The general linear hypothesis is true

Alternative Hypothesis:          The general linear hypothesis is not true

Data:                            Y

Sample Sizes:                    n1 = 43
                                 n2 = 14
                                 n3 = 21
                                 n4 = 72

Sample Dimension:                2000

Test Statistic:                  T[ZZ] = 121.1988

Approximation method to the      3-c matched chi^2-approximation
null distribution of T[ZZ]: 

Approximation parameter(s):      df    =   4.9334
                                 beta0 = -35.0606
                                 beta1 =   7.1068

P-value:                         0.0004959474
\end{CodeOutput}
\end{CodeChunk}

The second class of functions comprises  \code{FHW2004.GLHT.NABT()}, \code{SF2006.GLHT.NABT()}, 

\code{YS2012.GLHT.NABT()}, \code{ZZZ2022.GLHT.2cNRT()}, and \code{S2007.ks.NABT()}, all governed by the following five parameters:
\begin{itemize}
    \item \code{Y}\;	a list of $k$ data matrices. The $i$th element represents the data matrix ($n_i\times p$) from the $i$th population with each row representing a $p$-dimensional observation.
\item \code{X}\; a known $n\times f$ full-rank design matrix with $\operatorname{rank}(\mathbf{G})=f<n$.
\item \code{C}\; a known matrix of size $q\times f$ with $\operatorname{rank}(\mathbf{C})=q<f$.
\item \code{n}\;	 a vector of $k$ sample sizes. The $i$th element represents the sample size of group $i$, $n_i$. 
\item \code{p}\; the dimension of data.
\end{itemize}
To utilize these functions with the \proglang{corneal} dataset, we can directly employ the data matrix \code{corneal} as it contains all the subjects with each row representing a $p$-dimensional sample. We can set the matrices $\mathbf{X}$ and $\mathbf{C}$ corresponding to the one-way MANOVA problem as follows:
\begin{CodeChunk}
\begin{CodeInput}
R> q <- k-1
R> X <- matrix(c(rep(1,n[1]),rep(0,sum(n)),rep(1,n[2]), 
+    rep(0,sum(n)),rep(1,n[3]),rep(0,sum(n)),rep(1,n[4])),ncol=k,nrow=sum(n))
R> C <- cbind(diag(q),-rep(1,q))
\end{CodeInput}
\end{CodeChunk}

For instance, to apply \cite{Fujikoshi_2004}'s test, the corresponding code using 

\code{FHW2004.GLHT.NABT()} is:
\begin{CodeChunk}
\begin{CodeInput}
R> FHW2004.GLHT.NABT(Y,X,C,n,p) 
\end{CodeInput}
\begin{CodeOutput}
Results of Hypothesis Test
--------------------------

Test name:                       Fujikoshi et al. (2004)'s test

Null Hypothesis:                 The general linear hypothesis is true

Alternative Hypothesis:          The general linear hypothesis is not true

Data:                            Y

Sample Sizes:                    n1 = 43
                                 n2 = 14
                                 n3 = 21
                                 n4 = 72

Sample Dimension:                2000

Test Statistic:                  T[FHW] = 6.4015

Approximation method to the      Normal approximation
null distribution of T[FHW]: 

P-value:                         7.694084e-11
\end{CodeOutput}
\end{CodeChunk}
To conduct \cite{yamada2012test}'s test, we can use the code below:
\begin{CodeChunk}
\begin{CodeInput}
R> YS2012.GLHT.NABT(Y,X,C,n,p) 
\end{CodeInput}
\begin{CodeOutput}
Results of Hypothesis Test
--------------------------

Test name:                       Yamada and Srivastava (2012)'s test

Null Hypothesis:                 The general linear hypothesis is true

Alternative Hypothesis:          The general linear hypothesis is not true

Data:                            Y

Sample Sizes:                    n1 = 43
                                 n2 = 14
                                 n3 = 21
                                 n4 = 72

Sample Dimension:                2000

Test Statistic:                  T[YS] = 2.352

Approximation method to the      Normal approximation
null distribution of T[YS]: 

Approximation parameter(s):      Adjustment coefficient = 16.7845

P-value:                         0.009336667
\end{CodeOutput}
\end{CodeChunk}
The \code{cpn} is the adjustment coefficient used in the test.
For the normal-reference scale-invariant test, as proposed by \cite{Zhu_2023}, the corresponding code using \code{ZZZ2022.GLHT.2cNRT()} is:
\begin{CodeChunk}
\begin{CodeInput}
R> ZZZ2022.GLHT.2cNRT(Y,X,C,n,p) 
\end{CodeInput}
\begin{CodeOutput}
Results of Hypothesis Test
--------------------------

Test name:                       Zhu et al. (2022)'s test

Null Hypothesis:                 The general linear hypothesis is true

Alternative Hypothesis:          The general linear hypothesis is not true

Data:                            Y

Sample Sizes:                    n1 = 43
                                 n2 = 14
                                 n3 = 21
                                 n4 = 72

Sample Dimension:                2000

Test Statistic:                  T[ZZZ] = 5.5651

Approximation method to the      2-c matched chi^2-approximation
null distribution of T[ZZZ]: 

Approximation parameter(s):      df = 8.9706

P-value:                         1.083822e-07
\end{CodeOutput}
\end{CodeChunk}

The \proglang{R} code and test results of the rest tests for the GLHT problem are given in Appendix~\ref{app:glht}. 


To evaluate the performance of the ten GLHT methods on the \proglang{corneal} dataset, we examined whether the four corneal surface groups have the same mean corneal surface. The testing results are summarized in Table~\ref{tab:glht}. The results indicate that all ten tests strongly reject the null hypothesis, suggesting that the four groups are unlikely to have the same mean corneal surface.

Moreover, the estimated approximate degrees of freedom (d.f.) are relatively small, indicating that the underlying null distribution of the test statistics is likely skewed to the right. This skewness suggests that the normal approximation employed by some of the methods may not be adequate, and thus the reliability of their p-values may be compromised.




\begin{table}[!ht]
\centering
\begin{tabular}{lllll}
\toprule
\multicolumn{2}{c}{Approach}  &Function name& $p$-value & d.f. \\
\midrule \multirow{5}{*}{\shortstack[l]{NRTs}} & \multirow{3}{*}{2-c matched $\chi^2$-approx.} & \code{ZGZ2017.GLHT.2cNRT()} & $4.712\times 10^{-5}$ & 7.76 \\
\cmidrule{3-5} & & \code{ZZZ2022.GLHT.2cNRT()} & $1.084\times 10^{-7}$ & 8.97 \\
\cmidrule{3-5} & & \code{ZZG2022.GLHTBF.2cNRT()} & $2.577\times 10^{-4}$ & 6.17 \\
\cmidrule{2-5} & \multirow{2}{*}{3-c matched $\chi^2$-approx.} & \code{ZZ2022.GLHTBF.3cNRT()} & $4.959\times 10^{-4}$ & 4.93 \\
\cmidrule{3-5} & & \code{ZZ2022.GLHT.3cNRT()} & $9.145\times 10^{-5}$ & 6.05 \\
\midrule \multirow{5}{*}{\shortstack[l]{NABTs}} & & \code{FHW2004.GLHT.NABT()} & $7.694\times 10^{-11}$ & -- \\
\cmidrule{3-5} & & \code{SF2006.GLHT.NABT()} & $6.678\times 10^{-11}$ & -- \\
\cmidrule{3-5} & & \code{S2007.ks.NABT()} & $1.022\times 10^{-10}$ & -- \\
\cmidrule{3-5} & & \code{YS2012.GLHT.NABT()} & $9.337\times 10^{-3}$ & -- \\
\cmidrule{3-5} & & \code{ZGZ2017.GLHTBF.NABT()} & $1.177\times 10^{-10}$ & -- \\
\bottomrule
\end{tabular}
\caption{\label{tab:glht}  Testing results of one-way MANOVA for the corneal surface dataset using \pkg{HDNRA}.}
\end{table}

Given the high significance of the one-way MANOVA problem, the next point of interest is to examine whether there are differences in mean corneal surfaces between any two corneal groups. This can be achieved by straightforwardly adjusting the matrices $\mathbf{G}$ or $\mathbf{C}$. For instance, if our interest lies in testing whether the mean corneal surfaces differ between the normal group and the unilateral suspect group, we can set $\mathbf{G}=(1,-1,0,0)$ or $\mathbf{C}=(1,-1,0,0)$. This can be implemented using the following code:
\begin{CodeChunk}
\begin{CodeInput}
R> G <- t(as.matrix(c(1,-1, 0, 0)))
R> ZZG2022.GLHTBF.2cNRT(Y,G,n,p)
\end{CodeInput}
\begin{CodeOutput}
Results of Hypothesis Test
--------------------------

Test name:                       Zhang et al. (2022)'s test

Null Hypothesis:                 The general linear hypothesis is true

Alternative Hypothesis:          The general linear hypothesis is not true

Data:                            Y

Sample Sizes:                    n1 = 43
                                 n2 = 14
                                 n3 = 21
                                 n4 = 72

Sample Dimension:                2000

Test Statistic:                  T[ZZG] = 5.1536

Approximation method to the      2-c matched chi^2-approximation
null distribution of T[ZZG]: 

Approximation parameter(s):      df   = 2.1161
                                 beta = 7.5138

P-value:                         0.7353797
\end{CodeOutput}
\end{CodeChunk}

\begin{CodeChunk}
\begin{CodeInput}
R> C <- t(as.matrix(c(1,-1, 0, 0)))
R> ZZZ2022.GLHT.2cNRT(Y,X,C,n,p)
\end{CodeInput}
\begin{CodeOutput}
Results of Hypothesis Test
--------------------------

Test name:                       Zhu et al. (2022)'s test

Null Hypothesis:                 The general linear hypothesis is true

Alternative Hypothesis:          The general linear hypothesis is not true

Data:                            Y

Sample Sizes:                    n1 = 43
                                 n2 = 14
                                 n3 = 21
                                 n4 = 72

Sample Dimension:                2000

Test Statistic:                  T[ZZZ] = 0.6307

Approximation method to the      2-c matched chi^2-approximation
null distribution of T[ZZZ]: 

Approximation parameter(s):      df = 2.9902

P-value:                         0.5945916
\end{CodeOutput}
\end{CodeChunk}
Therefore, we cannot conclude that the mean corneal surfaces of the normal group and the unilateral suspect group are significantly different at the 5\% significance level. This conclusion is reasonable, given that the term "unilateral suspect" may indicate concerns about the health or characteristics of one eye, making it challenging to distinguish from the normal group.  Consequently, proposing strategies for early detection becomes imperative and essential in such cases.

\section{Comparison with other packages}\label{sec:5}



As noted in Section \ref{sec:intro}, there exist approximately 16 distinct packages pertaining to HDLSS equal-mean testing. Among these, only SKK-test is available in both \pkg{highDmean} and \pkg{HDNRA}, after excluding packages that cannot be installed. BS-test and SD-test are featured in both \pkg{SHT} and \pkg{HDNRA}, while BS-test, SD-test, and CQ-test are encompassed in both \pkg{highmean} and \pkg{HDNRA}.

In this section, we aim to compare the computational costs of identical tests from different packages, utilizing a variety of real datasets that span different values of dimension $p$ and total sample size $n$. The details of the datasets can be found at \url{https://github.com/nie23wp8738/i.i.d-high-dimensional-dataset}. 
By comparing these datasets across different $(p,n)$ regimes, we aim to assess the computational efficiency of our proposed package relative to other existing packages.

The execution time analysis for this section was performed on a MacBook Pro (16-inch, 2023, Apple M2 Max Chip, 32 GB RAM) running macOS Sequoia version 15.1. The integrated development environment (IDE) used was RStudio version 2024.09.0 Build 375 (\citealt{rstudio}), and computations were conducted with R version 4.4.1 (\citealt{Rversion}). Each test was executed 10 times, and the average execution time (in seconds) for these runs was recorded and presented in Tables~\ref{tab:dataset_comparison1} --\ref{tab:dataset_comparison3},  where ``NA" indicates that the package did not function correctly, while ``$>30$" signifies that no computational results were returned within 5 minutes. It is important to note that our package incorporates some efficient methods to reduce computational cost.  For example, when calculating $\tr(\hbSigma_i^2),i=1,2$, we express $ \hbSigma_i=(n_i-1)^{-1}\sum_{j=1}^{n_i} (\by_{ij}-\barby_i)(\by_{ij}-\barby_i)^{\top}=(n_i-1)^{-1}\bZ_i\bZ_i^\top, i=1,2$, where $\bZ_i=(\by_{i1}-\barby_i,\ldots,\by_{in_i}-\barby_i)$,  is a $ p\times n_i$ matrix. Utilizing the property $\tr(\bA^\top\bB)=\tr(\bA\bB^\top)$, we have
\[
\tr(\hbSigma_i^2)=(n_i-1)^{-2}\tr(\bZ_i\bZ_i^\top\bZ_i\bZ_i^\top)=(n_i-1)^{-2}\tr(\bZ_i^\top\bZ_i\bZ_i^\top\bZ_i), i=1,2.
\]
This allows us to adopt different strategies to minimize the computational cost:
\begin{itemize}
\item When  $n>p$, we employ $\tr(\hbSigma_i^2)=(n_i-1)^{-2}\tr(\bZ_i\bZ_i^\top\bZ_i\bZ_i^\top)$ with a computational complexity of  $\mathcal{O}(np^2)$. 
\item Conversely, when  $n<p$, we use $\tr(\hbSigma_i^2)=(n_i-1)^{-2}\tr(\bZ_i^\top\bZ_i\bZ_i^\top\bZ_i)$, resulting in a computational complexity of  $\mathcal{O}(n^2p)$ which is significantly smaller than   $\mathcal{O}(np^2)$.
\end{itemize}
As a result, the computational costs presented in Tables~\ref{tab:dataset_comparison1}--\ref{tab:dataset_comparison3} are ranked according to their respective computational complexities.

\begin{table}[!h]
\centering
\setlength{\tabcolsep}{1pt} 
\scalebox{0.75}{
\begin{tabular*}{1.3\textwidth}{@{\extracolsep{\fill}}llllll}
\toprule
 \multirow{2}{*}{Dataset} &\multirow{2}{*}{$(n,p)$} &\multicolumn{2}{l}{BS-test} &\multicolumn{2}{l}{SD-test}\\
 \cmidrule{3-4} \cmidrule{5-6}
& &  \pkg{SHT}  & \pkg{HDNRA}  &  \pkg{SHT}  & \pkg{HDNRA}\\
\midrule
  \proglang{SARS-CoV-2} &$(234,15979)$ & $>30$  & 0.4997 & $>30$ & 0.5352 \\
  \proglang{COVID-19} &$(86,20460)$ & $>30$  & 0.1358 & $>30$  & 0.1269 \\
  \proglang{Skeletal muscle} &$(36, 54675)$  & NA  & 0.0489 & NA  & 0.0585 \\
 \proglang{Pancreatic} &$(185,847)$ & 0.2861  & 0.0209 & 0.2950  & 0.0213 \\
  \proglang{Yeoh-V2} &$(94,2526)$ & 6.4349  & 0.0185 & 6.5678  & 0.0174 \\
  \proglang{Alizadeh-V3} &$(42, 2093)$ & 3.5892  & 0.0031  & 3.7291  & 0.0032 \\
 \proglang{Heart disease} &$(91,13)$& $9.2506 \times 10^{-5}$ & $3.8219 \times 10^{-5}$ & $1.1699 \times 10^{-4}$  & $4.0269 \times 10^{-5}$ \\
  \proglang{Rats} &$(20,17)$ & $11.2414\times 10^{-5}$  & $3.7169 \times 10^{-5}$ & $12.4526 \times 10^{-5}$  & $3.1495 \times 10^{-5}$ \\
\bottomrule
\end{tabular*}}
\caption{\label{tab:dataset_comparison1} Comparison of computational costs for \cite{bai1996effect}'s test (BS-test)  and \cite{Srivastava_2008}'s test (SD-test) from \pkg{SHT} and \pkg{HDNRA}.}
\end{table}

We first compare \pkg{SHT} and \pkg{HDNRA}, both of which include BS-test and SD-test, as shown in Table~\ref{tab:dataset_comparison1}. As expected, the execution time for functions in \pkg{HDNRA} decreases as their computational complexity is reduced. In contrast, the execution time for functions in \pkg{SHT} decreases primarily as the dimension $p$ decreases. This suggests that \pkg{SHT} does not implement the same optimization strategies as \pkg{HDNRA}, and instead follows a computational complexity of $\mathcal{O}(np^2)$ across both high-dimensional and low-dimensional datasets. Consequently, \pkg{SHT}  encounters significant computational challenges, particularly with  high-dimensional datasets such as the  \proglang{Skeletal muscle},  \proglang{COVID-19}, and  \proglang{SARS-CoV-2}. In contrast, \pkg{HDNRA} consistently outperforms \pkg{SHT} in computational efficiency across all high-dimensional datasets. 
For low-dimensional datasets, such as  \proglang{Rats} and  \proglang{Heart disease}, \pkg{SHT} exhibits comparable performance to \pkg{HDNRA}, with minimal differences in execution time between the two packages.

Next, we compare  \pkg{highmean} and \pkg{HDNRA}, both of which include BS-test, SD-test and CQ-test, as presented in Table~\ref{tab:dataset_comparison2}, and compare \pkg{highDmean} and \pkg{HDNRA}, both of which include SKK-test in Table~\ref{tab:dataset_comparison3}. We can get similar conclusions as those drawn from Table~\ref{tab:dataset_comparison1}. That is, \pkg{HDNRA} consistently demonstrates superior computational efficiency  than  \pkg{highmean} and \pkg{highDmean}  for high-dimensional datasets and comparable with its competitors  for low-dimensional datasets. 
\begin{table}[!h]
\centering
\setlength{\tabcolsep}{1pt} 
\scalebox{0.68}{
\begin{tabular*}{1.5\textwidth}{@{\extracolsep{\fill}}lllllllll}
\toprule
\multirow{2}{*}{Dataset} &\multirow{2}{*}{$(n,p)$} &\multicolumn{2}{l}{BS-test} &\multicolumn{2}{l}{SD-test} &\multicolumn{2}{l}{CQ-test}\\
 \cmidrule{3-4} \cmidrule{5-6} \cmidrule{7-8}
&&  \pkg{highmean}  & \pkg{HDNRA} &  \pkg{highmean}  & \pkg{HDNRA} & \pkg{highmean}  & \pkg{HDNRA} \\
\midrule
 \proglang{SARS-CoV-2} &$(234,15979)$ & 16.3799 & 0.4997 & 23.2747 & 0.5352 & 22.31 & 0.9760 \\
 \proglang{COVID-19} &$(86,20460)$  & 12.4355 & 0.1358 & 16.3887 & 0.1269  & 12.0107 & 0.1966 \\
\proglang{Skeletal muscle} &$(36, 54675)$ & NA & 0.0489  & NA & 0.0585 & NA & 0.0952 \\
 \proglang{Pancreatic} &$(185,847)$ & 0.0495 & 0.0209  & 0.0570 & 0.0213 & 0.0613 & 0.0319 \\
    \proglang{Yeoh-V2} &$(94,2526)$ & 0.2281 & 0.0185 & 0.2700 & 0.0174 & 0.2281 & 0.0270 \\
  \proglang{Alizadeh-V3} &$(42, 2093)$  & 0.0695 & 0.0031 & 0.1135 & 0.0032 & 0.0687 & 0.0046 \\
 \proglang{Heart disease} &$(91,13)$ & $5.2190 \times 10^{-5}$ & $3.8219 \times 10^{-5}$ & $6.2299 \times 10^{-5}$ & $4.0269 \times 10^{-5}$ & $1.2360 \times 10^{-4}$ & $2.6846 \times 10^{-4}$ \\  
  \proglang{Rats} &$(20,17)$  & $8.8716 \times 10^{-5}$ & $3.7169 \times 10^{-5}$ & $8.1720 \times 10^{-4}$ & $3.1495 \times 10^{-5}$ & $1.0650 \times 10^{-4}$ & $1.4942 \times 10^{-4}$ \\
\bottomrule
\end{tabular*}}
\caption{\label{tab:dataset_comparison2} Comparison of computational costs for \cite{bai1996effect}'s test (BS-test), \cite{Srivastava_2008}'s test (SD-test) and \cite{Chen_2010}'s test (CQ-test) from \pkg{highmean} and \pkg{HDNRA}.}
\end{table}
\begin{table}[!h]
\centering
\setlength{\tabcolsep}{1pt}
\scalebox{0.8}{
\begin{tabular*}{\textwidth}{@{\extracolsep{\fill}}llll}
\toprule
\multirow{2}{*}{Dataset} &\multirow{2}{*}{$(n,p)$} &\multicolumn{2}{l}{SKK-test} \\
\cmidrule{3-4}
& &  \pkg{highDmean}  & \pkg{HDNRA} \\
 \midrule
 \proglang{SARS-CoV-2} &$(234,15979)$  & $>30$  & 0.6144 \\
  \proglang{COVID-19} &$(86,20460)$  & $>30$  & 0.1541 \\
\proglang{Skeletal muscle} &$(36, 54675)$ & NA & 0.0697 \\
 \proglang{Pancreatic} &$(185,847)$ & 3.7722 & 0.0246 \\
  \proglang{Yeoh-V2} &$(94,2526)$  & $>30$  & 0.0213 \\
 \proglang{Alizadeh-V3} &$(42, 2093)$ & $>30$  & 0.0049 \\
\proglang{Heart disease} &$(91,13)$ & $5.0569 \times 10^{-4}$  & $6.6328\times 10^{-5}$ \\
 \proglang{Rats} &$(20,17)$ & $5.6369 \times 10^{-4}$  & $5.5695 \times 10^{-5}$ \\
\bottomrule
\end{tabular*}}
\caption{\label{tab:dataset_comparison3} Comparison of computational costs for \cite{Srivastava_2013}'s test (SKK-test) from \pkg{highDmean} and \pkg{HDNRA}.}
\end{table}


By delving into the code of the three competitors in detail, the aforementioned results can be explained as follows. First of all, the foundation of \pkg{HDNRA} relies on the efficient framework of \pkg{Rcpp} (\citealt{eddelbuettel2011rcpp}) and \pkg{RcppArmadillo} (\citealt{RcppArmadillo}) which not only facilitates seamless integration with \proglang{R} but also significantly accelerates the speed of the functions. Meanwhile, unlike \pkg{SHT}, there are almost no loops in our \proglang{C++} code. Therefore, even though the core of \pkg{SHT} is using \proglang{C++}, \pkg{SHT} still performs worse than \pkg{highmean} when running BS-test and SD-test based on the \proglang{Skeletal muscle}, \proglang{COVID-19}, and \proglang{SARS-CoV-2} datasets. Our package consistently outperforms the other three packages regardless of the dimension.

Secondly, our package addresses the presence of zero row vectors in the dataset by introducing a tiny constant ($10^{-10}$) to each $0$ as done in \pkg{highmean}. In some tests, such as SD-test and SKK-test, the existence of zero row vectors in the dataset will destroy the hope to find the inverse of some matrices, which will lead to the packages to fail. Conversely, \pkg{HDNRA} successfully overcame this challenge by adding a small positive constant ($10^{-10}$) to ensure numerical stability. This feature enables \pkg{HDNRA} to execute tests that would otherwise fail due to singular matrices, as evidenced by the results for the \proglang{Skeletal muscle} dataset.

\section{Summary and discussion}\label{sec:6}

In this paper, we initially present the theoretical attributes of the normal-reference approach and offer a synopsis of the related normal-reference tests. We highlight the \proglang{R} package \pkg{HDNRA}, showcasing its implementation not only for those normal-reference tests but also for various well-known tests addressing location testing problems for high-dimensional data. We conduct a comprehensive exploration of the normal-reference tests, highlighting their robust performance under mild conditions and ensuring effective size control. Both the two-sample problem and the general linear hypothesis testing (GLHT) problem can be addressed by this package which is a perspective not covered in the existing literature. Additionally, \pkg{HDNRA} incorporates two real datasets and features 21 tests, with their applications illustrated on the provided datasets. Furthermore, our package stands out by leveraging \proglang{C++} code throughout its framework, enabling quick and effective results in handling computationally intensive tasks when compared with some existing packages.

To the best of our knowledge, the current version of  \pkg{HDNRA} is the most comprehensive software tool for the normal-reference tests and some classical tests for high-dimensional data. Potential extensions and additional high-dimensional location tests may be incorporated in future updates to further enhance the capabilities of \pkg{HDNRA}.

\section*{Acknowledgments}\label{Acknow}
Wang and Zhu's work was supported by  the National Institute of Education, Singapore, under its Academic Research Fund (RI 4/22 ZTM) and Zhang's work  was supported by the National University of Singapore Academic Research grants (22-5699-A0001) and (23-1046-A0001).  The authors are grateful to Yehudit Hasin-Brumshtein (\url{yhasin@inflammatix.com}) for the permission to package the \proglang{COVID19} data to our \proglang{R} package.

\bibliography{refs}



 \begin{appendix}
 \section{R code: test results for the two-sample problem} \label{app:ts}
 In this section, we apply the two-sample tests included in the \pkg{HDNRA} package that were not covered in Section~\ref{2sfun.sec} to the \proglang{COVID19} dataset. The corresponding \proglang{R} code and test results are presented. 
 \begin{CodeChunk}
\begin{CodeInput}
R> data("COVID19")
R> group1 <- as.matrix(COVID19[c(2:19, 82:87), ])
R> group2 <- as.matrix(COVID19[-c(1:19, 82:87), ])
\end{CodeInput}
\end{CodeChunk}
For \cite{Zhang_2021}'s test, i.e., the normal-reference tests with 2-c matched $\chi^2$-approximation, the corresponding code of using \code{ZZGZ2021.TSBF.2cNRT()} is:
\begin{CodeChunk}
\begin{CodeInput}
R> ZZGZ2021.TSBF.2cNRT(group1,group2)
\end{CodeInput}
\begin{CodeOutput}
Results of Hypothesis Test
--------------------------

Test name:                       Zhang et al. (2021)'s test

Null Hypothesis:                 Difference between two mean vectors is 0

Alternative Hypothesis:          Difference between two mean vectors is not 0

Data:                            group1 and group2

Sample Sizes:                    n1 = 24
                                 n2 = 62

Sample Dimension:                20460

Test Statistic:                  T[ZZGZ] = 228972526332

Approximation method to the      2-c matched chi^2-approximation
null distribution of T[ZZGZ]: 

Approximation parameter(s):      df   = 2.73240e+00
                                 beta = 1.97808e+10

P-value:                         0.00693092
\end{CodeOutput}
\end{CodeChunk}
For the normal-reference scale-invariant test proposed by \cite{zhang2023two}, an additional parameter \code{cutoff} is required which is  an empirical criterion for applying the adjustment coefficient. The default value for \code{cutoff} is set to 1.2.
\begin{CodeChunk}
\begin{CodeInput} 
R> ZZZ2023.TSBF.2cNRT(group1,group2,cutoff = 1.2)
\end{CodeInput}
\begin{CodeOutput}
Results of Hypothesis Test
--------------------------

Test name:                       Zhang et al. (2023)'s test

Null Hypothesis:                 Difference between two mean vectors is 0

Alternative Hypothesis:          Difference between two mean vectors is not 0

Data:                            group1 and group2

Sample Sizes:                    n1 = 24
                                 n2 = 62

Sample Dimension:                20460

Test Statistic:                  T[ZZZ] = 6.511

Approximation method to the      2-c matched chi^2-approximation
null distribution of T[ZZZ]: 

Approximation parameter(s):      df  = 10.1280
                                 cpn = 17.9488

P-value:                         3.043196e-10
\end{CodeOutput}
\end{CodeChunk}

For \cite{zhang2022further}'s test, i.e., the normal-reference tests with 3-c matched $\chi^2$-approximation, the corresponding code of using \code{ZZ2022.TSBF.3cNRT()} is:
\begin{CodeChunk}
\begin{CodeInput}
R> ZZ2022.TSBF.3cNRT(group1,group2)
\end{CodeInput}
\begin{CodeOutput}
Results of Hypothesis Test
--------------------------

Test name:                       Zhang and Zhu (2022)'s test

Null Hypothesis:                 Difference between two mean vectors is 0

Alternative Hypothesis:          Difference between two mean vectors is not 0

Data:                            group1 and group2

Sample Sizes:                    n1 = 24
                                 n2 = 62

Sample Dimension:                20460

Test Statistic:                  T_ZZ = 10073497402

Approximation method to the      3-c matched chi^2-approximation
null distribution of T_ZZ: 

Approximation parameter(s):      df    =  1.912900e+00
                                 beta0 = -2.649726e+09
                                 beta1 =  1.385153e+09

P-value:                         0.009152364
\end{CodeOutput}
\end{CodeChunk}

 Presented below is the corresponding code for the two existing two-sample tests, namely, \cite{Srivastava_2008}'s and \cite{Chen_2010}'s tests.
\begin{CodeChunk}
\begin{CodeInput}
R> SD2008.TS.NABT(group1,group2)
\end{CodeInput}
\begin{CodeOutput}
Results of Hypothesis Test
--------------------------

Test name:                       Srivastava and Du (2008)'s test

Null Hypothesis:                 Difference between two mean vectors is 0

Alternative Hypothesis:          Difference between two mean vectors is not 0

Data:                            group1 and group2

Sample Sizes:                    n1 = 24
                                 n2 = 62

Sample Dimension:                20460

Test Statistic:                  T[SD] = 2.5704

Approximation method to the      Normal approximation
null distribution of T[SD]: 

Approximation parameter(s):      Adjustment coefficient = 15.2281

P-value:                         0.005078436
\end{CodeOutput}
\end{CodeChunk}
\begin{CodeChunk}
\begin{CodeInput}
R> CQ2010.TSBF.NABT(group1,group2)
\end{CodeInput}
\begin{CodeOutput}
Results of Hypothesis Test
--------------------------

Test name:                       Chen and Qin (2010)'s test

Null Hypothesis:                 Difference between two mean vectors is 0

Alternative Hypothesis:          Difference between two mean vectors is not 0

Data:                            group1 and group2

Sample Sizes:                    n1 = 24
                                 n2 = 62

Sample Dimension:                20460

Test Statistic:                  T[CQ] = 3.5355

Approximation method to the      Normal approximation
null distribution of T[CQ]: 

P-value:                         0.0002035166
\end{CodeOutput}
\end{CodeChunk}

\section{R code: obtaining empirical sizes in a two-sample problem} \label{app:ts_size}
Since the \proglang{R} packages \pkg{doParallel} (\citealt{doParallel}) combined with \pkg{foreach} (\citealt{foreach}) can enable the exploitation of multiple cores within a computing system to perform loops, thereby markedly enhancing computational efficiency. We calculate the empirical sizes displayed in Table~\ref{tab:ts_size} with the following code:
\begin{CodeChunk}
\begin{CodeInput}
### Helper function to extract p.value and parameters from NRtest object
R> extract_results <- function(result) {
### Check if the result is of class NRtest
+  if (inherits(result, "NRtest")) {
+    p.value <- result$p.value
+    parameters <- result$parameter
+    return(list(p.value = p.value, parameters = parameters))
+  } else {
+    return(list(p.value = NA, parameters = rep(NA, 2)))  
### Return NA if the result is not valid
+  }
+}
### Load required libraries
R> library(doParallel)
R> library(HDNRA)
R> library(foreach)
### Load the COVID19 dataset
R> data("COVID19")
R> group2 <- as.matrix(COVID19[-c(1:19, 82:87), ])
### Set the significance level
R> alpha <- 0.05
### Set up parallel computing environment
R> no_cores <- detectCores() - 1
R> cl <- makeCluster(no_cores)
R> registerDoParallel(cl)
### Number of repetitions for the analysis
R> nrep <- 10000
R> results <- foreach(N=1:nrep,.combine= 'rbind',.packages = c("HDNRA"))
+  %dopar%{
### Random sample for training data
+  train <- sample(1:62,31,replace = FALSE)
+  group2_1 <- group2[train,]
+  group2_2 <- group2[-train,]
### Statistical analysis
### Implement various statistical methods
+  ZGZC2020.TS.2cNRT <- extract_results(ZGZC2020.TS.2cNRT(group2_1, group2_2))
+  ZZZ2020.TS.2cNRT <- extract_results(ZZZ2020.TS.2cNRT(group2_1, group2_2))
+  ZZGZ2021.TSBF.2cNRT <- extract_results(ZZGZ2021.TSBF.2cNRT(group2_1, group2_2))
+  ZWZ2023.TSBF.2cNRT <- extract_results(ZWZ2023.TSBF.2cNRT(group2_1, group2_2))
+  ZZZ2023.TSBF.2cNRT <- extract_results(ZZZ2023.TSBF.2cNRT(group2_1, group2_2, 
                                        cutoff = 1.2))
+  ZZ2022.TS.3cNRT <- extract_results(ZZ2022.TS.3cNRT(group2_1, group2_2))
+  ZZ2022.TSBF.3cNRT <- extract_results(ZZ2022.TSBF.3cNRT(group2_1, group2_2))
+  BS1996.TS.NABT <- extract_results(BS1996.TS.NABT(group2_1, group2_2))
+  SD2008.TS.NABT <- extract_results(SD2008.TS.NABT(group2_1, group2_2))
+  CQ2010.TSBF.NABT <- extract_results(CQ2010.TSBF.NABT(group2_1, group2_2))
+  SKK2013.TSBF.NABT <- extract_results(SKK2013.TSBF.NABT(group2_1, group2_2))
### Convert p-values to binary (0 or 1) based on the significance level
+  size_ZGZC2020.TS.2cNRT <- as.integer(ZGZC2020.TS.2cNRT$p.value<alpha)
+  size_ZZZ2020.TS.2cNRT <- as.integer(ZZZ2020.TS.2cNRT$p.value<alpha)
+  size_ZZGZ2021.TSBF.2cNRT <- as.integer(ZZGZ2021.TSBF.2cNRT$p.value<alpha)
+  size_ZWZ2023.TSBF.2cNRT <- as.integer(ZWZ2023.TSBF.2cNRT$p.value<alpha)
+  size_ZZZ2023.TSBF.2cNRT <- as.integer(ZZZ2023.TSBF.2cNRT$p.value<alpha)
+  size_ZZ2022.TS.3cNRT <- as.integer(ZZ2022.TS.3cNRT$p.value<alpha)
+  size_ZZ2022.TSBF.3cNRT <- as.integer(ZZ2022.TSBF.3cNRT$p.value<alpha)
+  size_BS1996.TS.NABT <- as.integer(BS1996.TS.NABT$p.value<alpha)
+  size_SD2008.TS.NABT <- as.integer(SD2008.TS.NABT$p.value<alpha)
+  size_CQ2010.TSBF.NABT <- as.integer(CQ2010.TSBF.NABT$p.value<alpha)
+  size_SKK2013.TSBF.NABT <- as.integer(SKK2013.TSBF.NABT$p.value<alpha)
### Return combined results
+  return(c(size_ZGZC2020.TS.2cNRT, size_ZZZ2020.TS.2cNRT, size_ZZGZ2021.TSBF.2cNRT,
+           size_ZWZ2023.TSBF.2cNRT, size_ZZZ2023.TSBF.2cNRT, size_ZZ2022.TS.3cNRT, 
+           size_ZZ2022.TSBF.3cNRT, size_BS1996.TS.NABT, size_SD2008.TS.NABT, 
+           size_CQ2010.TSBF.NABT, size_SKK2013.TSBF.NABT,
+           ZGZC2020.TS.2cNRT$parameters[2], ZZZ2020.TS.2cNRT$parameters,
+           ZZGZ2021.TSBF.2cNRT$parameters[1], ZWZ2023.TSBF.2cNRT$parameters[1],
+           ZZZ2023.TSBF.2cNRT$parameters[1], ZZZ2023.TSBF.2cNRT$parameters[2],
+           ZZ2022.TS.3cNRT$parameters[3], ZZ2022.TSBF.3cNRT$parameters[1],
+           SD2008.TS.NABT$parameters, SKK2013.TSBF.NABT$parameters))
+}
R> stopCluster(cl)
R> results <- as.matrix(results)
R> size <- as.matrix(results[,1:11])
R> para <- as.matrix(results[,12:21])
R> print(colSums(size)/nrep)
R> print(colSums(para)/nrep)
\end{CodeInput}
\end{CodeChunk}
\section{R code: test results for the GLHT problem} \label{app:glht}
 In this section, we apply the tests for the GLHT problem included in the \pkg{HDNRA} package that were not covered in Section~\ref{glhtfun.sec} to the \proglang{corneal} dataset. The corresponding \proglang{R} code and test results are presented. 
\begin{CodeChunk}
\begin{CodeInput}
R> data("corneal")
R> p <- dim(corneal)[2]
R> k <- 4
R> Y <- list()
R> group1 <- as.matrix(corneal[1:43, ])
R> group2 <- as.matrix(corneal[44:57, ])
R> group3 <- as.matrix(corneal[58:78, ])
R> group4 <- as.matrix(corneal[79:150, ])
R> Y[[1]] <- group1
R> Y[[2]] <- group2
R> Y[[3]] <- group3
R> Y[[4]] <- group4
R> n <- c(ncol(Y[[1]]),ncol(Y[[2]]),ncol(Y[[3]]),ncol(Y[[4]]))
R> G <- cbind(diag(k-1),rep(-1,k-1))
R> q <- k-1
R> X <- matrix(c(rep(1,n[1]),rep(0,sum(n)),rep(1,n[2]), 
+    rep(0,sum(n)),rep(1,n[3]),rep(0,sum(n)),rep(1,n[4])),ncol=k,nrow=sum(n))
R> C <- cbind(diag(q),-rep(1,q))
\end{CodeInput}
\end{CodeChunk}

For \cite{Zhang_2017}'s test, which is a normal-reference test with a 2-c matched $\chi^2$-approximation, the corresponding code using \code{ZGZ2017.GLHT.2cNRT()} is:
\begin{CodeChunk}
\begin{CodeInput}
R> ZGZ2017.GLHT.2cNRT(Y,G,n,p)
\end{CodeInput}
\begin{CodeOutput}
Results of Hypothesis Test
--------------------------

Test name:                       Zhang et al. (2017)'s test

Null Hypothesis:                 The general linear hypothesis is true

Alternative Hypothesis:          The general linear hypothesis is not true

Data:                            Y

Sample Sizes:                    n1 = 43
                                 n2 = 14
                                 n3 = 21
                                 n4 = 72

Sample Dimension:                2000

Test Statistic:                  T[ZGZ] = 159.7325

Approximation method to the      2-c matched chi^2-approximation
null distribution of T[ZGZ]: 

Approximation parameter(s):      df   = 7.7601
                                 beta = 4.8153

P-value:                         4.711944e-05
\end{CodeOutput}
\end{CodeChunk}
To implement \cite{zhang2022revisit}'s test, a normal-reference test with a 3-c matched $\chi^2$-approximation, we can use the following code with \code{ZZ2022.GLHT.3cNRT()}:
\begin{CodeChunk}
\begin{CodeInput}
R> ZZ2022.GLHT.3cNRT(Y,G,n,p)
\end{CodeInput}
\begin{CodeOutput}
Results of Hypothesis Test
--------------------------

Test name:                       Zhu and Zhang (2022)'s test

Null Hypothesis:                 The general linear hypothesis is true

Alternative Hypothesis:          The general linear hypothesis is not true

Data:                            Y

Sample Sizes:                    n1 = 43
                                 n2 = 14
                                 n3 = 21
                                 n4 = 72

Sample Dimension:                2000

Test Statistic:                  T[ZZ] = 122.1679

Approximation method to the      3-c matched chi^2-approximation
null distribution of T[ZZ]: 

Approximation parameter(s):      df    =   6.0490
                                 beta0 = -33.4163
                                 beta1 =   5.5243

P-value:                         9.144893e-05
\end{CodeOutput}
\end{CodeChunk}

Presented below are the corresponding code for the other existing tests for the GLHT problem, namely, the tests proposed by \cite{Srivastava_2006}, \cite{Schott_2007}, and \cite{Zhou_2017}, respectively.
\begin{CodeChunk}
\begin{CodeInput}
R> SF2006.GLHT.NABT(Y,X,C,n,p)
\end{CodeInput}
\begin{CodeOutput}
Results of Hypothesis Test
--------------------------

Test name:                       Srivastava and Fujikoshi (2006)'s test 

Null Hypothesis:                 The general linear hypothesis is true

Alternative Hypothesis:          The general linear hypothesis is not true

Data:                            Y

Sample Sizes:                    n1 = 43
                                 n2 = 14
                                 n3 = 21
                                 n4 = 72

Sample Dimension:                2000

Test Statistic:                  T[SF] = 6.4231

Approximation method to the      Normal approximation
null distribution of T[SF]: 

P-value:                         6.677982e-11
\end{CodeOutput}
\end{CodeChunk}
\begin{CodeChunk}
\begin{CodeInput}
R> S2007.ks.NABT(Y,n,p)
\end{CodeInput}
\begin{CodeOutput}
Results of Hypothesis Test
--------------------------

Test name:                       Schott (2007)'s test

Null Hypothesis:                 Difference between k mean vectors is 0

Alternative Hypothesis:          Difference between k mean vectors is not 0

Data:                            Y

Sample Sizes:                    n1 = 43
                                 n2 = 14
                                 n3 = 21
                                 n4 = 72

Sample Dimension:                2000

Test Statistic:                  T[S] = 6.3581

Approximation method to the      Normal approximation
null distribution of T[S]: 

P-value:                         1.021509e-10
\end{CodeOutput}
\end{CodeChunk}
\begin{CodeChunk}
\begin{CodeInput}
R> ZGZ2017.GLHTBF.NABT(Y,G,n,p)
\end{CodeInput}
\begin{CodeOutput}
Results of Hypothesis Test
--------------------------

Test name:                       Zhou et al. (2017)'s test

Null Hypothesis:                 The general linear hypothesis is true

Alternative Hypothesis:          The general linear hypothesis is not true

Data:                            Y

Sample Sizes:                    n1 = 43
                                 n2 = 14
                                 n3 = 21
                                 n4 = 72

Sample Dimension:                2000

Test Statistic:                  T[ZGZ] = 121.1988

Approximation method to the      Normal approximation
null distribution of T[ZGZ]: 

P-value:                         1.176941e-10
\end{CodeOutput}
\end{CodeChunk}

\section{R code: assessing computational costs across different packages} \label{app:time}
The following code is the core part employed to evaluate the computational costs presented in Tables~\ref{tab:dataset_comparison1}--\ref{tab:dataset_comparison3}. The rationale is to calculate the time difference between \code{starttime} and \code{endtime} as a measure of the computational costs of the targeted function. You can check the \code{time.R} file in \url{https://github.com/nie23wp8738/i.i.d-high-dimensional-dataset} to obtain the complete code or the complete code demonstrated on the 8 datasets for this part.

\begin{CodeChunk}
\begin{CodeInput}
# Load the necessary libraries for hypothesis testing
R> library(HDNRA)
R> library(highmean)
R> library(highDmean)
R> library(SHT)
\end{CodeInput}
\end{CodeChunk}
\begin{CodeChunk}
\begin{CodeInput}
# Function to run and time a test 10 times
# Helper function to compute the average time taken by each test over 10 repetitions
R> run_test_10_times <- function(test_function, group1, group2) {
+  times <- replicate(10, {
+    startTime <- Sys.time()
+    test_function(group1, group2)
+    endTime <- Sys.time()
+    as.numeric(difftime(endTime, startTime, units = "secs"))
+  })
+  mean(times)  # Return the average time
+}
\end{CodeInput}
\end{CodeChunk}

\begin{CodeChunk}
\begin{CodeInput}
## HDNRA and highDmean comparison
# SKK_test (high-dimensional two-sample test for mean differences)
R> average_time_SKK_test <- run_test_10_times(SKK_test, group1, group2)
R> cat("Average time for SKK_test: ", average_time_SKK_test, " seconds\n")

# SKK2013.TSBF.NABT (high-dimensional two-sample test based on Srivastava, Katayama, 
and Kano's method)
R> average_time_SKK2013 <- run_test_10_times(SKK2013.TSBF.NABT, group1, group2)
R> cat("Average time for SKK2013.TSBF.NABT: ", average_time_SKK2013, " seconds\n")
\end{CodeInput}
\end{CodeChunk}

\begin{CodeChunk}
\begin{CodeInput}
## HDNRA and SHT comparison
# mean2.1996BS (Bai and Saranadasa's mean test)
R> average_time_mean2_1996BS <- run_test_10_times(mean2.1996BS, group1, group2)
R> cat("Average time for mean2.1996BS: ", average_time_mean2_1996BS, " seconds\n")

# BS1996.TS.NABT (Test statistic based on Bai and Saranadasa)
R> average_time_BS1996_TS_NABT <- run_test_10_times(BS1996.TS.NABT, group1, group2)
R> cat("Average time for BS1996.TS.NABT: ", average_time_BS1996_TS_NABT, " seconds\n")

# mean2.2008SD (Srivastava and Du's mean test)
R> average_time_mean2_2008SD <- run_test_10_times(mean2.2008SD, group1, group2)
R> cat("Average time for mean2.2008SD: ", average_time_mean2_2008SD, " seconds\n")

# SD2008.TS.NABT (Srivastava and Du's test statistic)
R> average_time_SD2008_TS_NABT <- run_test_10_times(SD2008.TS.NABT, group1, group2)
R> cat("Average time for SD2008.TS.NABT: ", average_time_SD2008_TS_NABT, " seconds\n")
\end{CodeInput}
\end{CodeChunk}

\begin{CodeChunk}
\begin{CodeInput}
## HDNRA and highmean comparison
# apval_Bai1996 (Adjusted p-value using Bai's method)
R> average_time_apval_Bai1996 <- run_test_10_times(apval_Bai1996, group1, group2)
R> cat("Average time for apval_Bai1996: ", average_time_apval_Bai1996, " seconds\n")

# BS1996.TS.NABT (Test statistic based on Bai and Saranadasa)
R> average_time_BS1996_TS_NART <- run_test_10_times(BS1996.TS.NABT, group1, group2)
R> cat("Average time for BS1996.TS.NABT: ", average_time_BS1996_TS_NABT, " seconds\n")

# apval_Sri2008 (Adjusted p-value using Srivastava's method)
R> average_time_apval_Sri2008 <- run_test_10_times(apval_Sri2008, group1, group2)
R> cat("Average time for apval_Sri2008: ", average_time_apval_Sri2008, " seconds\n")

# SD2008.TS.NABT (Srivastava and Du's test statistic)
R> average_time_SD2008_TS_NABT <- run_test_10_times(SD2008.TS.NABT, group1, group2)
R> cat("Average time for SD2008.TS.NABT: ", average_time_SD2008_TS_NABT, " seconds\n")

# apval_Chen2010 (Chen and Qin's mean test for equal covariance)
R> run_test_10_times_Chen2010 <- function(group1, group2) {
+  times <- replicate(10, {
+    startTime <- Sys.time()
+    apval_Chen2010(group1, group2, eq.cov = TRUE)
+    endTime <- Sys.time()
+    as.numeric(difftime(endTime, startTime, units = "secs"))
+  })
+  mean(times)  # Return the average time
+}

R> average_time_apval_Chen2010 <- run_test_10_times_Chen2010(group1, group2)
R> cat("Average time for apval_Chen2010: ", average_time_apval_Chen2010, " seconds\n")

# CQ2010.TSBF.NABT (Test statistic based on Chen and Qin's method)
R> average_time_CQ2010_TSBF_NABT <- run_test_10_times(CQ2010.TSBF.NABT, group1, group2)
R> cat("Average time for CQ2010.TSBF.NABT: ", average_time_CQ2010_TSBF_NABT, " seconds\n")
\end{CodeInput}
\end{CodeChunk}

 \end{appendix}


\end{document}